\documentclass[12pt]{iopart}

\usepackage{iopams}  
\usepackage{cite}
\usepackage{hyperref}  
\usepackage{wasysym}
\usepackage{graphicx}
\usepackage{lineno}

\begin{document}

\newcommand{\peii}{$\pi_{e2}$}
\newcommand{\peiig}{$\pi_{e2\gamma}$}
\newcommand{\peiiee}{$\pi_{e2ee}$}
\newcommand{\peiii}{$\pi_{e3}$}
\newcommand{\VmA}{$V$$-$$A$}  


\title[Pion rare decay measurements]{
          Experimental study of rare charged pion decays}

\author{Dinko Po\v{c}ani\'c$^1$, Emil Frle\v{z}$^1$ and
         Andries van der Schaaf$^{\,2}$}

\address{$^1$ 
  Institute for Nuclear and Particle Physics and Department of Physics, \\
  \hspace*{5pt} University of Virginia, Charlottesville, VA 22904-4714, USA } 

\address{$^2$ Physik-Institut, Universit\"at Z\"urich, CH-8057 Z\"urich,
  Switzerland} 

\ead{\mailto{pocanic@virginia.edu}}

\begin{abstract}

The combination of simple dynamics, small number of available decay
channels, and extremely well controlled radiative and loop corrections,
make charged pion decays a sensitive means for testing the underlying
symmetries and the universality of weak fermion couplings, as well as
for improving our understanding of pion structure and chiral dynamics.
This paper reviews the current state of experimental study of the
allowed rare decays of charged pions: (a) leptonic, $\pi^+ \to
e^+\nu_e$, or $\pi_{e2}$, (b) radiative, $\pi^+ \to e^+\nu_e\gamma$, or
$\pi_{e2\gamma}$, and $\pi^+ \to e^+\nu_e e^+e^-$, or $\pi_{e2ee}$, and
(c) semileptonic, $\pi^+\to \pi^0 e^+ \nu$, or $\pi_{e3}$.  Taken
together, the combined data set presents an internally consistent
picture that also agrees well with standard model predictions.  The
internal consistency is illustrated well by the $\pi_{e2}$ branching
ratio of $(R_{e/\mu}^\pi)^{\rm PIBETA} = (1.2366 \pm 0.0064) \times
10^{-4}$ extracted in this work from the PIBETA measurement of the
$\pi_{e3}$ decay and the current best value for the CKM matrix element
$V_{ud}$.  However, even after the great progress of the recent decades,
experimental precision is lagging far behind that of the theoretical
description for all above processes.  We review the implications of the
present state of knowledge and prospects for further improvement in the
near term.

\end{abstract}

\submitto{\jpg}
\maketitle

\section{Introduction}

Pi mesons hold a special place in both the weak and the strong
interactions, and remain subjects of research interest ever since their
discovery almost 70 years ago \cite{Lat47}.

Historically, pion decay has provided an important testing ground for
the weak interaction and radiative corrections from the earliest time of
the development of modern particle theory.  Decays of the charged pion
proceed via the weak interaction, and therefore closely reflect its
properties and dynamics.  In particular, the suppression of the direct
electronic decay of the pion ($\pi\to e\nu$, or \peii), manifested in
the failure of initial searches to observe it, led to an early
examination of the nature of the weak interaction governing pion
decay \cite{Rud49}.  A low branching fraction of $\sim 1.3\times
10^{-4}$ was predicted \cite{Fey58} even before the decay's
discovery \cite{Faz59}, as a direct consequence of the \VmA\ nature of
the weak interaction, through helicity suppression of the right-handed
state of the electron.  Not only was this prediction confirmed by the
early measurements \cite{Faz59,Imp59,And59}, the size of the radiative
corrections \cite{Ber58,Kin59} was soon proved correct as
well \cite{And60,DiC64}.

In more recent times pion decays have been theoretically described with
extraordinary precision.  Thanks to the underlying symmetries and the
associated conservation laws, the more complicated and thus more
uncertain hadronic processes are suppressed.  If measured with precision
comparable to that of their theoretical description, pion decays offer
an outstanding, clean testing ground of universality of lepton and quark
couplings.  Any documented deviation from standard model expectations
would indicate presence of dynamics not included in the SM, affecting
pion decays through loop diagrams.

In the hadronic sector the pion plays the role of the long-range
exchange particle in effective nucleon-nucleon Lagrangians.  Just as
importantly, the pion also plays the role of the pseudo-Goldstone boson
in the breaking of chiral symmetry for hadrons \cite{Gol61,Gol62}.
Whereas early formulations of the phenomenon were focusing on nucleons
as the fundamental hadronic fermions, the concept remains valid in the
quark picture.  While a massless pion would suffice to account for the
spontaneous chiral symmetry breaking, the nonzero pion mass
\cite{Nam61,Gel68} can be related directly to the light quark masses
within the framework of the explicit chiral symmetry breaking,
particularly in chiral perturbation theory (ChPT)
\cite{Wei79,Gas82,Gas84,Pic95}.  Although in principle the full theory
of quantum chromodynamics (QCD) contains all of this physics, in
practice QCD does not readily provide manageable calculations in the low
energy, nonperturbative domain.  The situation will improve in the near
future as lattice QCD calculations continue to increase accuracy,
reliability and reach.  Thus, for a long time chiral perturbation theory
and other similar effective Lagrangian approximate methods have provided
the primary and the only practical systematic treatment of low energy
hadronic processes based on QCD.  Basic properties of the pion, such as
its charge radius and form factors, are closely related to the constants
of the chiral Lagrangian \cite{Don92} and are thus fundamental to
nonperturbative QCD calculations.

In the remainder of this work we review experimental studies of
individual rare decay modes of the charged pion in the order of
descending branching fraction, with a focus on recent and ongoing
experiments.  It is worth noting that almost all measurements to date
have studied the positive pion.  Stopped $\pi^-$ mesons are captured
with high probability in target nuclei (and undergo strong
interactions), leaving few, if any, pions to decay weakly.  Hence, in
practice $\pi^-$ decays can only be studied in flight in vacuo, with
typically a tiny fraction of the beam pions decaying in the apparatus,
and the remaining pions escaping before decaying.  Positive pions can be
easily stopped in a designated target, with a small fraction lost
through prompt hadronic reactions with matter required to stop the beam.
The stopped $\pi^+$'s decay with an exponential time distribution
governed by the 26\,ns pion lifetime.  The dominant decay mode is
$\pi^+\to\mu^+\nu_\mu$ ($\pi_{\mu2}$) which will not be discussed here.
The resulting 4.1\,MeV muons are typically contained in the target and
decay via the $\mu^+\to e^+\nu_e\bar{\nu}_\mu$ process with
$\tau_\mu \simeq 2.2\,\mu$s.

\section{\texorpdfstring
    {\boldmath The electronic $\pi\to e\bar{\nu}$ decay (\peii)}
    {The electronic pi e nu decay} \label{sec:pi_e2} }

\subsection{Motivation: processes affecting electron-muon universality}

The $\pi^- \to \ell\bar{\nu}_\ell$ (or, equivalently,
$\pi^+\to \bar{\ell}\nu_\ell$) decay connects a pseudoscalar $0^-$ state
(the pion) to the vacuum.  At the lowest, tree level, the ratio of the
$\pi \to e\bar{\nu}$ to $\pi \to \mu\bar{\nu}$ decay widths is given by
\cite{Fey58,Bry82}
\begin{equation}
    R_{e/\mu,0}^\pi = \frac{\Gamma(\pi \to  e\bar{\nu})}
          {\Gamma(\pi \to  \mu\bar{\nu})}
       = \frac{m_e^2}{m_\mu^2}\cdot
        \frac{(m_\pi^2-m_e^2)^2}{(m_\pi^2-m_\mu^2)^2}
      \simeq 1.283 \times 10^{-4}\,.  \label{eq:pi_e2_tree}
\end{equation}
The first factor in the above expression, the ratio of squared lepton
masses for the two decays, comes from the helicity suppression by the
\VmA\ lepton-$W$ boson weak couplings.  If, instead, the decay could
proceed directly through the pseudoscalar current, the ratio
$R_{e/\mu}^\pi$ would reduce to the second, phase-space factor, or
approximately 5.5.  More complete treatment of the process includes
$\delta R_{e/\mu}^\pi$, the radiative and loop corrections, and the
possibility of lepton universality violation, i.e., that $g_e$ and
$g_\mu$, the electron and muon couplings to the $W$, respectively, may
not be equal:
\begin{equation}
    R_{e/\mu}^\pi =\frac{\Gamma(\pi \to  e\bar{\nu}(\gamma))}
          {\Gamma(\pi \to  \mu\bar{\nu}(\gamma))}
      = \frac{g_e^2}{g_\mu^2}\frac{m_e^2}{m_\mu^2}
        \frac{(m_\pi^2-m_e^2)^2}{(m_\pi^2-m_\mu^2)^2}
           \left(1+\delta R_{e/\mu}^\pi \right)\,,
    \label{eq:pi_e2_general}
\end{equation}
where the ``$(\gamma)$'' indicates that radiative decays are fully
included in the branching fractions.  Steady improvements of the
theoretical description of the $\pi_{e2}$ decay since the 1950's have
recently culminated in a series of calculations that have refined the
Standard Model (SM) prediction to a precision of 8 parts in $10^5$:
\begin{equation}
     \left(R_{e/\mu}^{\pi}\right)^{\rm SM} = 
       \frac{\Gamma(\pi \to e\bar{\nu}(\gamma))}
          {\Gamma(\pi \to  \mu\bar{\nu}(\gamma))}\bigg|_{\rm {calc}} =
   \cases{
    1.2352(5) \times 10^{-4} & {\rm Ref.~\cite{Mar93},} \\
    1.2354(2) \times 10^{-4} & {\rm Ref.~\cite{Fin96},} \\
    1.2352(1) \times 10^{-4} & {\rm Ref.~\cite{Cir07}.} 
         }  \label{eq:pi_e2_full_SM}
\end{equation}
A comparison with equation \eref{eq:pi_e2_tree} reveals that the
radiative and loop corrections amount to almost 4\% of
$R_{e/\mu}^{\pi}$.  However, as discussed below, the experimental
precision at this time lags behind the theoretical one by more than an
order of magnitude.

Because of the large helicity suppression of the \peii\ decay, its
branching ratio is highly susceptible to small non-\VmA\ contributions
from new physics, making this decay a particularly suitable subject of
study, as discussed in, e.g.,
Refs.~\citen{Sch81,Sha82,Loi04,Ram07,Cam05,Cam08}.  This prospect
provides the primary motivation for the ongoing PEN\cite{PENweb} and
PiENu\cite{PiENuWeb} experiments.  Of the possible ``new physics''
contributions in the Lagrangian, \peii\ is directly sensitive to the
pseudoscalar one.  At the precision of $10^{-3}$, $R_{e/\mu}^\pi$ probes
the pseudoscalar and axial vector mass scales up to 1,000\,TeV and
20\,TeV, respectively\cite{Cam05,Cam08}.  For comparison,
Cabibbo-Kobayashi-Maskawa (CKM) matrix unitarity and precise
measurements of several superallowed nuclear beta decays constrain the
non-SM vector contributions to $>20\,$TeV, and scalar to $>10\,$TeV
\cite{PDG12}.  Although scalar interactions do not directly contribute
to $R_{e/\mu}^\pi$, they can do so through loop diagrams, resulting in
sensitivity to new scalar interactions up to 60\,TeV \cite{Cam05,Cam08}.
The subject was recently reviewed at length in Ref.~\cite{Bry11}.  In
addition, $(R_{e/\mu}^{\pi})^{\rm exp}$ provides limits on masses of
certain SUSY partners\cite{Ram07}, and on neutrino sector
anomalies\cite{Loi04}.

\subsection{\texorpdfstring
  {Past measurements of the $\pi\to e\bar{\nu}$ decay branching ratio}
  {Past measurements of the pi to e nu decay branching fraction}}

Even though $R_{e/\mu}^{\pi}$ is small, of order $10^{-4}$, it would be
relatively straightforward to measure precisely because of the large
difference in decay ejectile energies, 69.79\,MeV vs. 4.12\,MeV for the
$e$ and $\mu$ channels, respectively.  However, the subsequent decay of
the muon, $\mu\to e \bar{\nu}_e{\nu}_\mu$, with the endpoint energy of
52.83\,MeV, effectively brings the two processes kinematically closer,
making the clean identification of the direct $\pi \to e$ events
challenging in a ``tail'' overlap region discussed below.  This
complication is partly mitigated by the $\sim 85$ times longer lifetime
of the muon than that of the pion.  Selecting only early decays, say,
within a couple of pion lifetimes, suppresses the sequential
$\pi\to\mu\to e$ events by more than an order of magnitude, thus
effectively enhancing the apparent $\pi\to e\bar{\nu}$ branching ratio.
Nearly all experiments to date have used a variant of the stopped pion
decay technique to measure $R_{e/\mu}^{\pi}$.

The technique based on detecting and normalizing $\pi\to e$ to
$\pi\to\mu\to e$ decays at rest in a nonmagnetic spectrometer is
inherently independent of several important sources of systematic
uncertainty that are shared by the two processes, and that consequently
cancel in the ratio.  Prime examples of shared uncertainties that cancel
include the relative placement of the target and the positron detector
and the resulting acceptance.  (On the other hand, using a magnetic
spectrometer removes this insensitivity, as it would introduce a strong
dependence of the solid angle on $e^+$ momentum.)  Further cancellations
occur for the efficiency of positron detection (largely independent of
$E_{e^+}$, although corrections at the $10^{-3}$--$10^{-2}$ level are
typically needed), various beam properties, efficiency of pion stop
identification in the target, overall trigger efficiency etc.  All of
these quantities would have to be determined with higher precision in an
experiment that would normalize the detected \peii\ decay events to the
corresponding number of beam pions stopped in the target.  Additional
quantities such as the time displacement of the first detection interval
from the arrival time of the pion and duration of the detection time
gate, affect the extraction of $R_{e/\mu}^{\pi}$ in well understood
ways.

The first notable measurement of $R_{e/\mu}^{\pi}$ with 6\% uncertainty
(rather than an upper limit or detection of a small number of \peii\
decay events), was performed by the University of Chicago group with
positive pions stopped inside a double-focusing magnetic
spectrometer \cite{And60}.  This work was soon followed by a $\sim 2$\%
measurement \cite{DiC64} at the Columbia Nevis synchrocyclotron
laboratory; decay positrons from $\pi^+$'s stopped in an active target
were detected in a \diameter 23\,cm\,$\times$\,24\,cm NaI(Tl) detector.
Given the NaI detector size and acceptance for photons, the analysis
included low-$E_\gamma$ inner bremsstrahlung (IB) radiative $\pi^+\to
e^+\nu_e\gamma$ events in the branching fraction, in accordance with the
theoretical definition of $R_{e/\mu}^{\pi}$, as given
in \eref{eq:pi_e2_general}.

The Nevis measurement brought to light an important experimental
uncertainty, related to how accurately the NaI(Tl) energy response
function is known.  Positrons in the few to a hundred MeV energy range
generate electromagnetic showers that cannot be completely contained in
a finite-sized physical detector volume; there is a non-zero probability
that some energy will escape the volume, primarily in the form of
photons, inducing a low-energy ``tail'' in the response function to a
monoenergetic positron.  (Use of a magnetic spectrometer would eliminate
this problem, though at the cost of significantly more complicated
acceptance systematics.)  This intrinsic instrumental tail coexists with
the physical ``tail'' of radiative decay events for which the
accompanying photon escaped detection, primarily due to limited detector
solid angle.  We note in passing that in a nearly hermetic detector,
radiative decays will generate a high energy ``tail'' extending above
the $\sim$\,69.8\,MeV two-body decay positron energy; this was not a
feature of the early measurements.  In a sufficiently segmented detector
such events can be properly treated by analyzing $(E+pc)/2$ rather than
$E$ alone.  

The tail correction for $\pi\to e\nu$ events falling below the $\mu$
decay endpoint energy in the Nevis experiment was 9.1\% with a
$\sim$\,0.7\% combined systematic and statistical uncertainty.  While
this correction diminishes for larger detectors, especially with
external active enclosures, it remains a dominant source of systematic
uncertainty for $R_{e/\mu}^{\pi}$, especially as experiments approach
the theoretical precision of equation \eref{eq:pi_e2_full_SM}.

The TRIUMF group of Bryman and collaborators \cite{Bry86} used a similar
technique to that used in the Nevis measurement, with improvements.
TINA, the TRIUMF NaI(Tl) detector was larger at \diameter 46\,cm
$\times$ 51\,cm long, and subtended a solid angle of $\simeq 0.7$\% of
$4\pi$\,sr, with energy resolution of about 4\% full width at
half-maximum (FWHM) at 70\,MeV, almost four times better than the Nevis
detector's $\sim$\,15\%.  The experiment collected approximately $3.2
\times 10^4$ \peii\ decay events.  The resulting branching ratio had a
1.2\% overall uncertainty.  The single largest contribution to the
uncertainty, at 0.75\%, arose from the tail correction.

\begin{figure}[hbt]
  \hspace*{0.15\linewidth}  
  \includegraphics[width=0.6\linewidth]{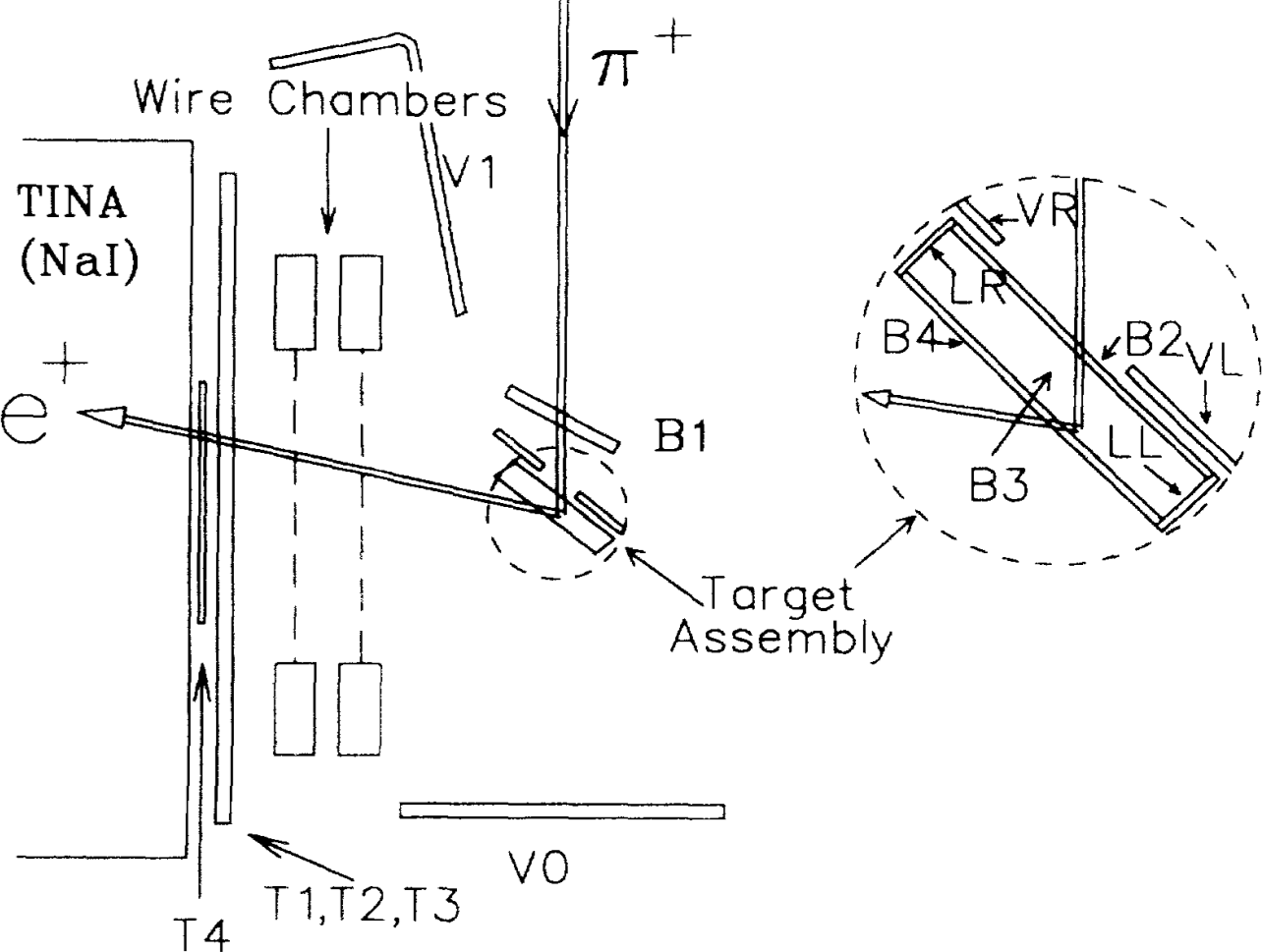} 
  \caption{TRIUMF TINA experimental setup, with beam (B), veto (V),
    trigger (T) counters, and beam entering from top of page.  For
    further details see \cite{Bri92}. } 
   \label{fig:TRIUMF_92}
\end{figure}

TINA was used again in a follow-up measurement (\fref{fig:TRIUMF_92})
by the same group \cite{Bri92}, with numerous improvements in
experimental design including an increased solid angle coverage of 2.9\%
of $4\pi$\,sr and an order of magnitude more detected \peii\ decay
events.  The measured branching fraction
\begin{equation}
   R_{e/\mu}^\pi =
    [1. 2265 \pm 0.0034(\textrm{stat}) \pm 0.0044(\textrm{syst})]
      \times 10^{-4} \,, \label{eq:pi_e2_britt_92}
\end{equation}
is in excellent agreement with standard model predictions.

\begin{figure}[htb]
  \hspace*{0.15\linewidth}
  \includegraphics[width=0.6\linewidth]{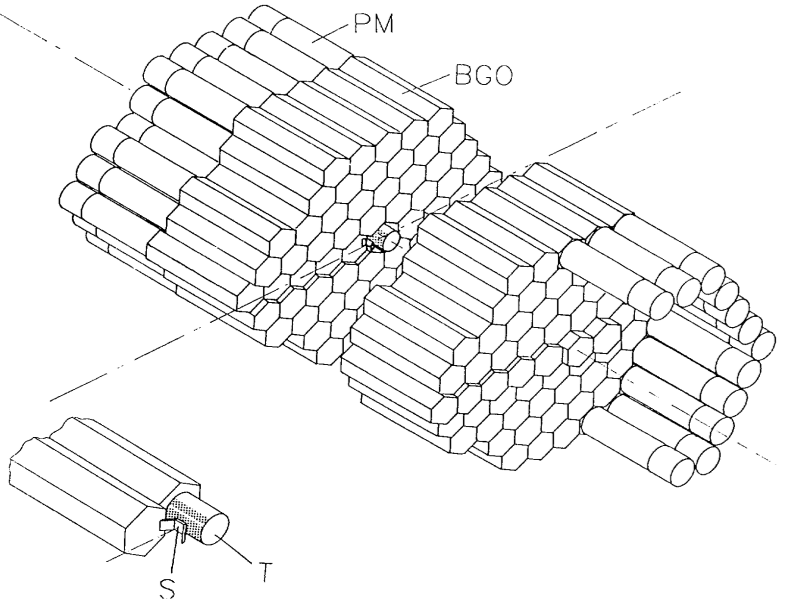} 
  \caption{Bern-PSI BGO calorimeter apparatus, shown with the two
    BGO ``walls'' separated, and the closeup of the last beam defining
    scintillator (S) and active target (T) counters; the pion beam enters
    from the left.  For further details see \cite{Cza93}. }
   \label{fig:Bern-PSI_93}
\end{figure}

At the Paul Scherrer Institute (PSI) a Bern--PSI group performed a
measurement of $R_{e/\mu}^{\pi}$ at a similar level of precision using a
radically different apparatus \cite{Cza93}, essentially simultaneously
with the TRIUMF measurement.  The Bern--PSI setup is shown
in \fref{fig:Bern-PSI_93}.  The key component of the
apparatus was a nearly hermetic $\sim 4\pi$\,sr calorimeter consisting
of 132 bismuth germanium oxide (BGO) detectors in the shape of hexagonal
prisms, each 20\,cm or 18 radiation lengths long, and with inscribed
radius of 5.5\,cm.  The detectors were arranged so that a narrow opening
allowed the beam (almost equal parts pions and muons) to stop in a small
cylindrical plastic scintillator (active target), whose light was read
out by a central calorimeter crystal.  The resulting rms resolution was
sub-2\% at 70\,MeV.  Because of the hermetic nature of the calorimeter,
radiative muon decays presented a significant background under the
$\pi\to e\bar{\nu}(\gamma)$ signal peak in the energy spectrum.  A total
of $3\times 10^5$ $\pi\to e$, and $1.2\times 10^6$ $\pi\to \mu$ events
were recorded.  The resulting branching fraction was
\[
   R_{e/\mu}^\pi =
    [1. 2346 \pm 0.0035(\textrm{stat}) \pm 0.0036(\textrm{syst})]
      \times 10^{-4} \,, \label{eq:pi_e2_bern-PSI_93}  
\]
with the systematic uncertainty coming primarily from corrections for
photonuclear reactions, radiative $\mu$ decay background, and
electromagnetic losses.

\Tref{tab:pe2_BRs} summarizes the available experimental results
on $R_{e/\mu}^{\pi}$. 
\begin{table}[tb]
  \caption{\label{tab:pe2_BRs}Summary of published experimental results
  on $R_{e/\mu}^{\pi}$, and the Particle Data Group (PDG) average.
                 }
  \begin{indented}
    \item[]
      \begin{tabular}{@{}lccc}
        \br
        Experiment & \peii\ events & $10^4\cdot R_{e/\mu}^{\pi}$  & Reference \\
        \mr
        U. Chicago/EFINS&  1.2k &  $1.21\pm 0.07$ & \cite{And60}\\
        Columbia/Nevis  &  10.8k & $1.247\pm 0.028$& \cite{DiC64}\\
        TRIUMF/TINA     &   32k  & $1.218\pm 0.014$ & \cite{Bry86}\\
        TRIUMF/TINA     &  190k  & $1.2265\pm 0.0056$& \cite{Bri92}\\
        Bern--PSI       &  120k  & $1.2346\pm 0.0050$& \cite{Cza93}\\
        \mr
        PDG average   &  342k  & $1.230 \pm 0.004$ & \cite{PDG12}\\
        \br
      \end{tabular}
    \end{indented}
\end{table}
Combined, the 1992/93 TRIUMF and PSI results define the present
experimental precision for the decay \cite{PDG12}, as the earlier
measurements do not significantly affect the current global average:
\begin{equation}
   \left(R_{e/\mu}^{\pi}\right)^{\rm exp} = 
       (1.230 \pm 0.004) \times 10^{-4}\,. \label{eq:pi_e2_avg}
\end{equation}
This experimental world average lags behind the theoretical precision
of \eref{eq:pi_e2_full_SM} by more than an order of magnitude.  The
precision gap, already notable in 1993, has only increased with the
passage of time, and has motivated a new generation of experiments
currently under way, and briefly discussed below.

\subsection{The PEN experiment at PSI \label{sec:pen}}

\begin{figure}[tb]
    \includegraphics[width=\linewidth]{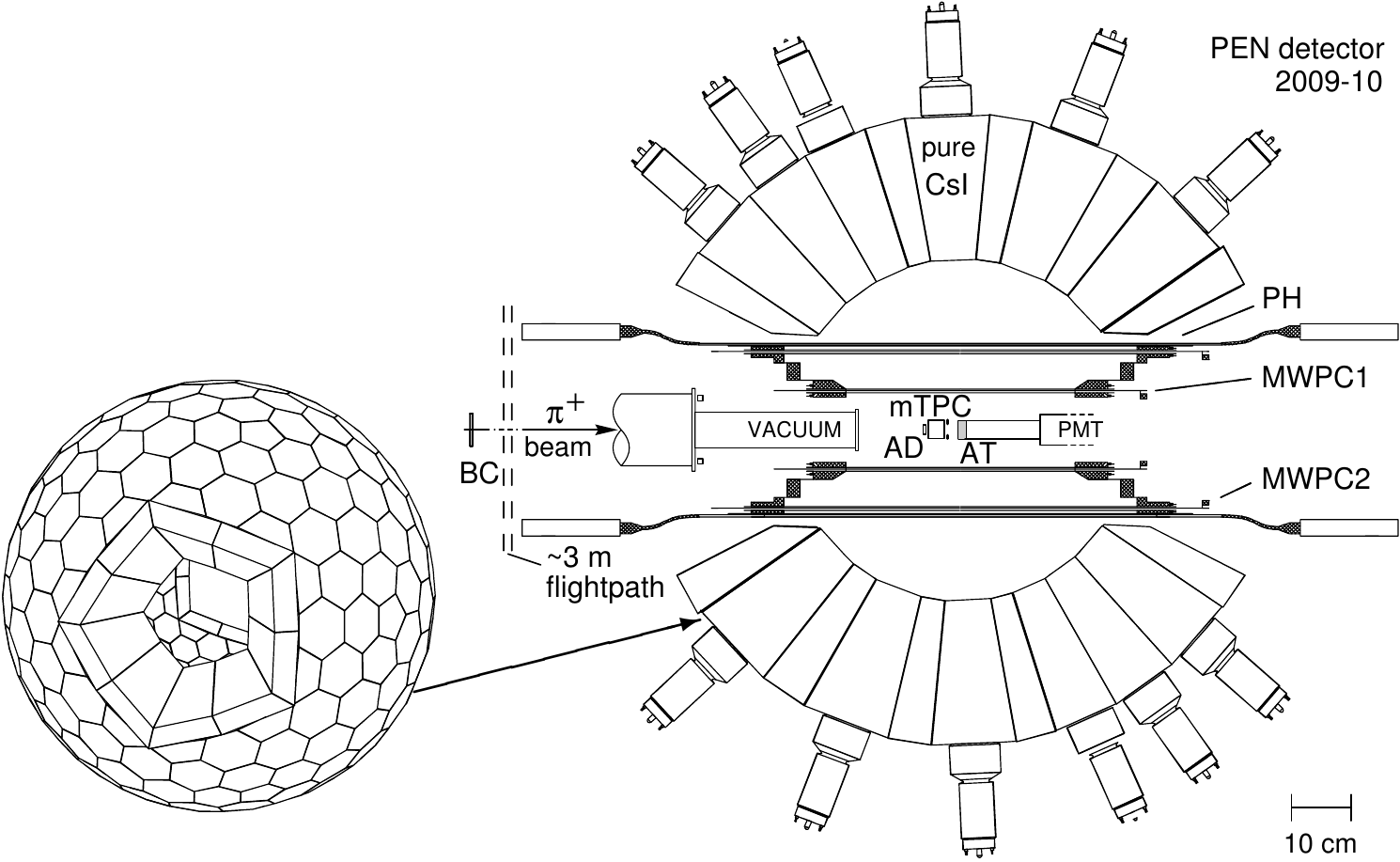}
    \caption{Schematic cross section of the PIBETA/PEN apparatus, shown
      in the 2009 PEN configuration, with its main components: beam
      entry with the upstream beam counter (BC), 5\,mm thick active
      degrader (AD), mini time projection chamber (mTPC) followed by a
      passive Al collimator, and active target (AT), cylindrical
      multiwire proportional chambers (MWPC's), plastic hodoscope (PH)
      detectors and photomultiplier tubes (PMT's), 240-element pure CsI
      electromagnetic shower calorimeter and its PMT's.  BC, AD, AT and
      PH detectors are made of plastic scintillator.  For details
      concerning the detector performance see \cite{Frl04a}.}
      \label{fig:PEN_det}
\end{figure}

In 2006 a new measurement of $R_{e/\mu}^{\pi}$ was proposed at the Paul
Scherrer Institute by a collaboration of seven institutions from the US
and Europe \cite{PENweb}, with the aim to reach
\begin{equation}
   \frac{\Delta R_{e/\mu}^{\pi}}{R_{e/\mu}^{\pi}}
       \simeq 5 \times 10^{-4} \,.    \label{eq:pen_goal}
\end{equation}
The PEN experiment uses an upgraded version of the PIBETA detector
system, described in detail in Ref.~\cite{Frl04a}, and previously used
in a series of rare pion and muon decay measurements
\cite{Poc04,Frl04b,Byc09,Poc14}.  The main component of the PEN
apparatus, shown in \fref{fig:PEN_det}, 
is a spherical large-acceptance ($\sim\,3\pi$\,sr) electromagnetic
shower calorimeter.  The calorimeter consists of 240 truncated hexagonal
and pentagonal pyramids of pure CsI, 22\,cm or 12 radiation lengths
deep.  The inner and outer diameters of the sphere are 52\,cm and
96\,cm, respectively.  Beam particles entering the apparatus with
$p\simeq 75$\,MeV/$c$ are first tagged in a thin upstream beam counter
(BC) and refocused by a triplet of quadrupole magnets.  After a $\sim
3$\,m long flight path they pass through a 5\,mm thick active degrader
(AD) and a low-mass mini time projection chamber (mTPC), finally to
reach a 15\,mm thick active target (AT) where the beam pions stop.
Decay particles are tracked non-magnetically in a pair of concentric
cylindrical multiwire proportional chambers (MWPC1,2) and an array of
twenty 4\,mm thick plastic hodoscope detectors (PH), all surrounding the
active target.  The BC, AD, AT and PH detectors are all made of fast
plastic scintillator material and read out by fast photomultiplier tubes
(PMTs).  Signals from the beam detectors are sent to waveform
digitizers, running at 2\,GS/s for BC, AD, and AT, and at 250\,MS/s for
the mTPC.

\begin{figure}[b]
  \hspace*{\fill}
  \parbox{0.48\linewidth}{
          \includegraphics[width=\linewidth]{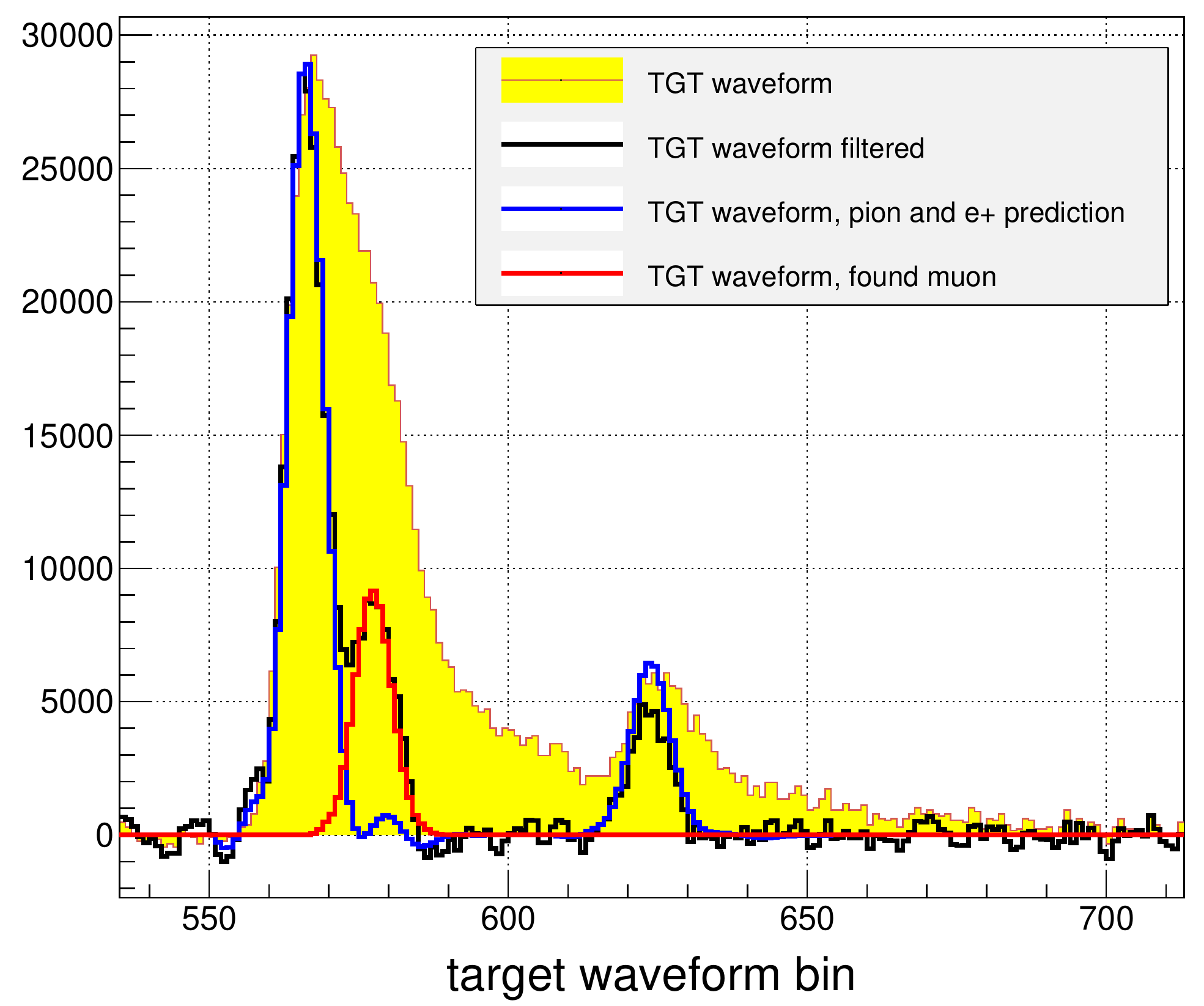}
                         }
  \parbox{0.48\linewidth}{
          \includegraphics[width=\linewidth]{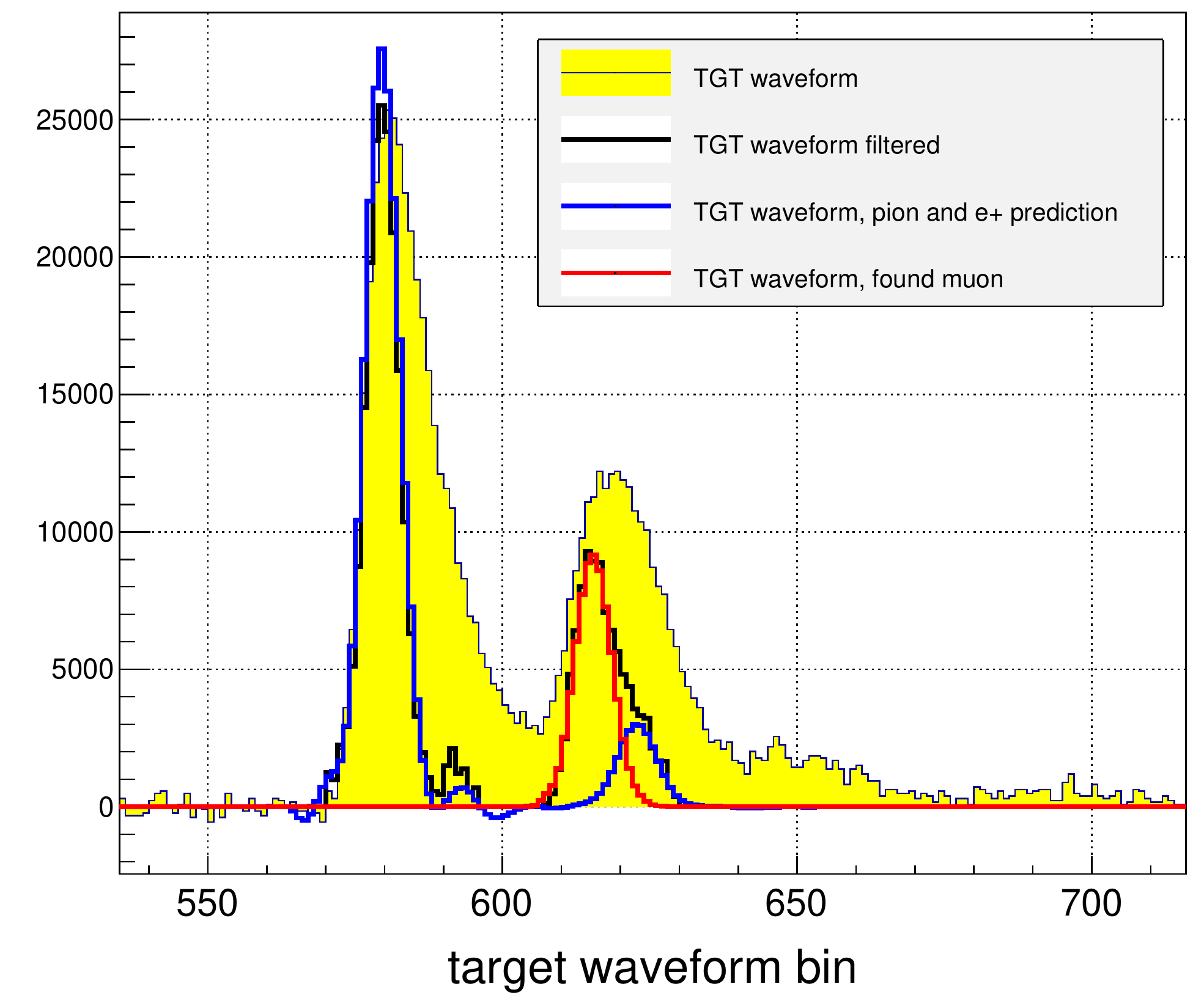}
                         } 

  \caption{Full and filtered active target (TGT) waveform in the PEN
    experiment for two challenging $\pi\to\mu\to e$ sequential decay
    events with an early $\pi\to\mu$ decay (left) and early $\mu\to e$
    decay (right).  The filtering procedure consists of a simple
    algebraic manipulation of the signal.  To the naked eye both raw
    waveforms appear to have two peaks only.  The separation of events
    with/without a muon signal depends critically on the accuracy of the
    predictions for the pion and positron signals.  For the pion the
    prediction is based on the times and energies observed in BC and
    AD.  For the positron the prediction depends on PH timing and the
    pathlength reconstructed with the pion and positron tracking
    detectors (see \fref{fig:tgt_chsq}).}  \label{fig:wf_fits}
\end{figure}

\begin{figure}[htb]
   \hspace*{\fill}
    \includegraphics[width=\linewidth]{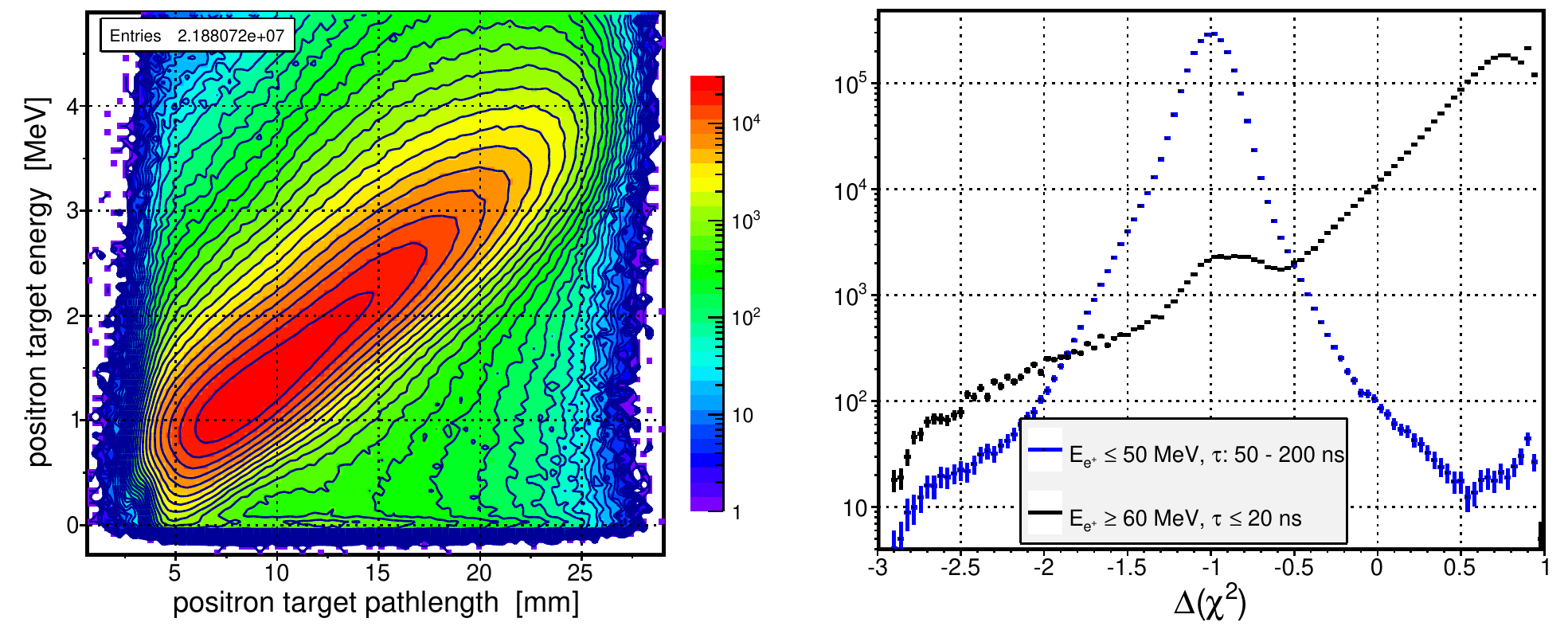}
   \caption{Left: correlation between observed positron energy in the
     target waveform and the $e^+$ path length in the target,
     reconstructed from the observed $\pi^+$ and $e^+$ trajectories.
     Shown are events with proper $\pi\to\mu\to e$ sequences for which
     the $e^+$ signal is well separated from other signals.  Right:
     difference in $\chi^2$ for the assumptions of a target waveform
     with/without a muon pulse present.  The observable is normalized
     such that $\pi\to e\nu$ events peak at $+1$, and $\pi\to\mu\to e$
     at $-1$.  Shown are events for two different combinations of $e^+$
     energy and decay time resulting in almost pure samples of $\pi\to
     e\nu$ and $\pi\to\mu\to e$, respectively.  Tiny admixtures of 
     another process are readily identified and are of great help in
     reducing the systematic uncertainties. } \label{fig:tgt_chsq}
\end{figure}

As discussed in the preceding section, a key source of systematic
uncertainty in previous \peii\ experiments has been the hard to measure
low energy tail of the detector response function, caused by
electromagnetic shower leakage from the calorimeter, mostly in the form
of photons.  PEN is no exception in this respect.  Other physical
processes, if not properly identified and suppressed, also contribute
events to the low energy part of the spectrum; unlike shower leakage
they can also produce high energy events.  One process is ordinary pion
decay into a muon in flight, before the pion is stopped, with the
resulting muon decaying within the time gate accepted in the
measurement.  Another is the unavoidable physical process of radiative
decay.  The latter is well measured and properly accounted for in the
PEN apparatus, as discussed below in the section on radiative decay.
Shower leakage and pion decays in flight can only be well characterized
if the $\pi\to\mu\to e$ chain can be well separated from the direct
$\pi \to e$ decay in the target.  Therefore much effort has been devoted
to digitization, filtering and analysis of the target
waveforms \cite{Pal12}, as illustrated in \fref{fig:wf_fits}.
The method used for separating the 2-peak (\peii) and 3-peak
($\pi\to\mu\to e$) events is illustrated and explained
in \fref{fig:tgt_chsq}.
The key to the method is provided by the beam and MWPC detectors which
are used to predict the pion and positron energy deposition in the
target, and the times of their signals.  Once the predicted waveform is
subtracted the net waveform is scanned for the presence of a 4.1\,MeV
muon peak.  The difference between the minimum $\chi^2$ values with and
without the muon peak is reported as $\Delta(\chi^2)$, constructed so
that clean 2-peak and 3-peak fits return values of $+1$ and $-1$,
respectively.  The scan is fast and returns a $\Delta(\chi^2)$ value for
every event, as illustrated in the figure.

\begin{figure}[t]
  \hspace*{0.1\linewidth}
   \includegraphics[width=0.9\linewidth]{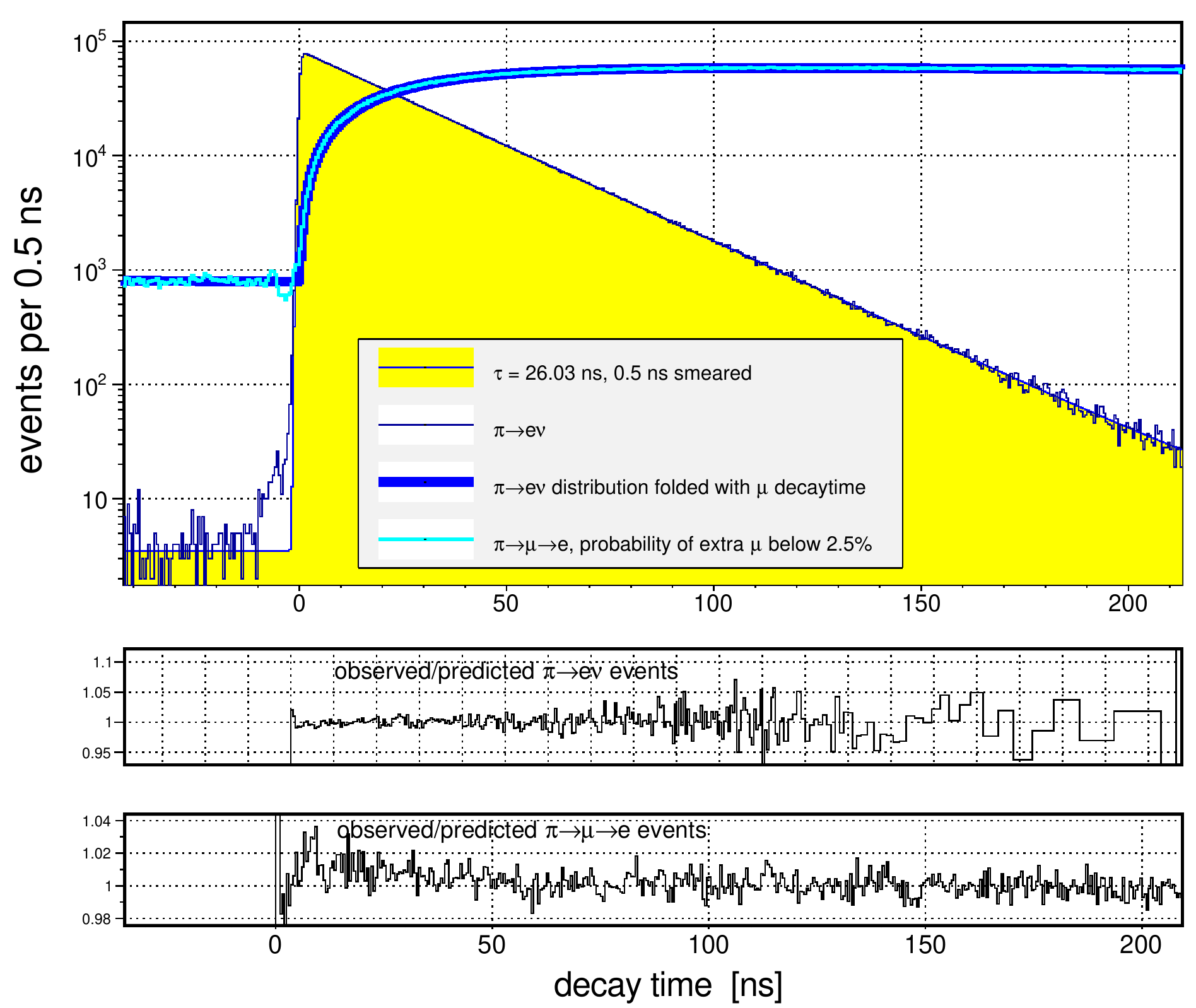}
  \caption{Decay time histograms for a subset of 2010 PEN data: $\pi\to
    e\nu$ events and $\pi\to\mu\to e$ sequential decay events.  The two
    processes are distinguished primarily by the total $e^+$ energy and
    by the absence or presence, respectively, of an extra 4.1\,MeV
    (muon) in the target due to $\pi\to\mu$ decay.  The $\pi_{e2}$ data
    are shown with a pion lifetime $\tau_\pi=26.03$\,ns exponential
    decay function superimposed.  The $\pi\to\mu\to e$ data were
    prescaled by a factor of $\sim$\,1/64; they are shown with the cut
    on the probability of $< 2.5$\% for a second, pile-up muon to be
    present in the target at $t=0$, the time of the nominal pion stop.
    The turquoise histogram gives the $\pi\to\mu\to e$ yield constructed
    entirely from the measured $\pi\to e\nu$ data folded with the $\mu$
    decay rate, and corrected for random muons; it perfectly matches the
    bold dark blue histogram.  The two lower plots show the observed to
    predicted ratios for \peii\ and $\pi\to\mu\to e$ events,
    respectively.  The scatter in the ratio plots is statistical in
    nature. } \label{fig:pen-decaytime}
\end{figure}

A particularly telling figure regarding the PEN data quality are the
decay time histograms of the $\pi\to e\nu$ decay and $\pi\to \mu\to e$
sequence, shown in \fref{fig:pen-decaytime} for a subset of data
recorded in 2010.
The $\pi\to e\nu$ data follow the exponential decay law over more than
three orders of magnitude, and perfectly predict the measured
$\pi\to\mu\to e$ sequential decay data once the latter are corrected for
random (pile-up) events.  Both event ensembles were obtained with
minimal requirements (cuts) on detector observables, none of which bias
the selection in ways that would affect the branching ratio.  The
probability of random $\mu\to e$ events originating in the target can be
controlled in the data sample by making use of multihit time to digital
converter (TDC) data that record early pion stop signals.  With this
information one can strongly suppress events in which an ``old'' muon
was present in the target by the time of the pion stop that triggered
the readout.

The ``intrinsic'' low energy tail of the PEN response function below
$\sim$ 50\,MeV, due to shower losses for \peii\ decay events for pions
at rest, amounts to approximately 2\% of the full yield.  Events with
$\pi\to\mu$ decays in flight, with subsequent ordinary Michel decay of
the stopped muon in the target, add a comparable contribution to the
tail.  The two contributions can be simulated accurately, with the
respective detector responses independently verified through comparisons
with measured data in appropriately selected processes and regions of
phase space.  These response functions are entered into the maximum
likelihood analysis (MLA), used to describe all measured processes
simultaneously.  The quantity $R_{e/\mu}^\pi$ is evaluated in the MLA as
the ratio of the magnitudes found for the $\pi\to e(\gamma)$ and
$\pi\to\mu\to e$ processes.  Although verified through comparisons with
Monte Carlo simulations, the intrinsic tail itself is not directly
measurable at the required precision because of the statistical
uncertainties arising in the tail data selection procedure.  Radiative
decay processes are directly measurable and accounted for in the MLA
procedure.  More information about the PEN/PIBETA detector response
functions is given below, in the section on \peiii\ decays, and in
\cite{Frl04a}.

During the 2008-10 production runs the PEN experiment accumulated some
$2.3 \times 10^7$ $\pi\to e\nu$, and more than $1.5 \times 10^8$
$\pi\to\mu\to e$ events, as well as significant numbers of pion and muon
radiative decays.  A comprehensive blinded maximum likelihood analysis
is under way to extract a new experimental value of $R_{e/\mu}^{\pi}$.
As of this writing, there appear no obstacles that would prevent the PEN
collaboration to reach a precision of $\Delta R/R < 10^{-3}$.  The
competing PiENu experiment at TRIUMF, discussed below, has a similar
precision goal.  The near to medium future will thus bring about a
substantial improvement in the limits on $e$-$\mu$ lepton universality,
and the attendant SM limits.

\subsection{The PiENu experiment at TRIUMF}

The new PIENU experiment at TRIUMF builds on the earlier measurements at
the same laboratory \cite{Bri92}, aiming at a significant improvement in
precision through refinements of the technique used.  The branching
fraction will again be obtained from the ratio of positron yields from
the $\pi\to e\nu$ decay and from the $\pi\to\mu\to e$ decay chain.  As
in other experiments that detect decay positrons in a nonmagnetic
spectrometer, many normalization factors, such as the solid angle of
positron detection, cancel to first order in PiENu, leaving only
corrections for small energy-dependent effects.  Major improvements in
precision in PiENu over the earlier TRIUMF TINA measurement derive from
improved geometry and beamline, a superior calorimeter, as well as
high-speed digitizing of all detector signals.

\begin{figure}[tbh]
  \hspace*{0.15\linewidth}
     \includegraphics[width=0.7\linewidth]{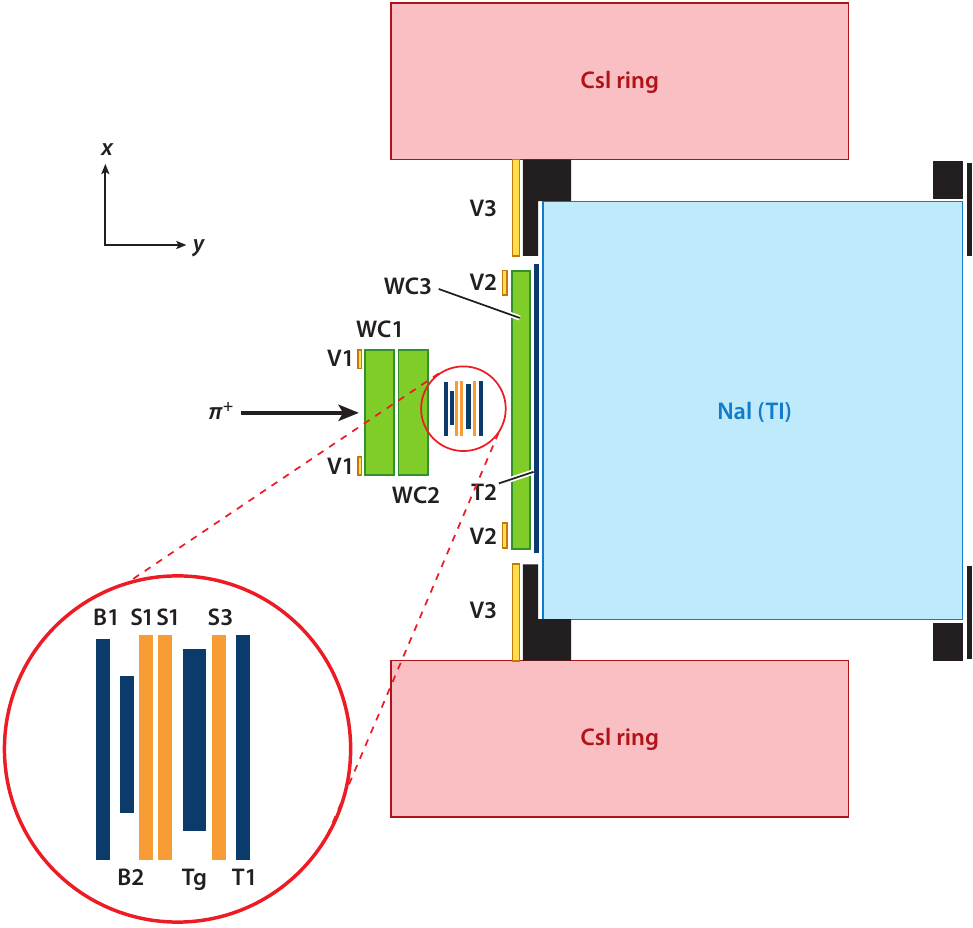}
  \caption{The TRIUMF PiENu experiment setup \cite{PiENuWeb}.  The
    beam enters from the left. ``B'' and ``T'' refer to the beam pion
    and positron telescope plastic scintillation counters,
    respectively.  For more details see
    \cite{PiENuWeb,Bry11,Agu09,Agu10}.} 
    \label{fig:PiENu_app}
\end{figure}

\Fref{fig:PiENu_app} shows a schematic rendition of the PiENu
experimental apparatus.  A 75\,MeV/$c$ $\pi^+$ beam from the improved
TRIUMF M13 beam line \cite{Agu09} is tracked in wire chambers,
identified by plastic scintillators, and stopped in a 0.8\,cm thick
scintillator target.  Fine tracking near the target is provided by two
sets of single-sided silicon strip detectors located immediately
upstream and downstream of the target assembly.  The positrons from
$\pi\to e\nu$ and $\pi\to\mu\to e$ decays are detected in the positron
telescope, which consists of a silicon strip counter, two thin plastic
counters, and an acceptance-defining wire chamber that covers the front
of the crystal calorimeter.  The solid-angle acceptance of the telescope
counters is 20\% of 4$\pi$\,sr.  

The primary energy measurement is performed using BINA, a cylindrical,
single-crystal \diameter 48\,cm $\times$ 48\,cm long NaI(Tl) detector
\cite{Bla96}.  BINA's energy resolution of approximately 2.2\% FWHM for
incident positrons at 70\,MeV \cite{Agu10} is better by a factor of
approximately two than the resolution observed in the previous
measurement with TINA \cite{Bri92}.  The NaI detector is surrounded by
two layers consisting of 97 pure CsI crystals, 8.5\,cm thick and
2$\times$25\,cm long \cite{Chi95,Kom98} to capture electromagnetic
shower leakage from BINA, thus helping to suppress the instrumental
low-energy ``tail''.

As in PEN, analog signals from plastic scintillators, silicon strip,
NaI, and CsI detectors are recorded as waveforms, using appropriately
fast digitizers.  To suppress background arising from old muon decay
signals in the target region and to reduce possible distortions in the
time spectrum due to pileup, the incident pion rate is kept at $6\times
10^4$/s.

The PiENu collaboration has accumulated upwards of $2 \times 10^7$
$\pi\to e\nu$ events through 2012.  Combining the corresponding
statistical uncertainty with reduced systematic uncertainties, the
collaboration expects to reach $\Delta R_{e/\mu}^{\pi}/R_{e/\mu}^{\pi} <
0.1$\%.  The data acquisition phase of the experiment indeed ended in
2012, and, as of this writing, the collaboration is working on data
analysis.

\section{\texorpdfstring
    {\boldmath Radiative decays $\pi\to e\nu\gamma$ (\peiig) and
      $\pi\to e\nu\gamma e^+e^-$ (\peiiee)}
    {Radiative decays  pi to e nu gamma (pi-e2g) and pi to e nu e+e-
      decay} \label{sec:rpd}  } 

\subsection{General considerations}

\begin{figure}[tb]
  \centerline{\includegraphics[width=0.7\linewidth]{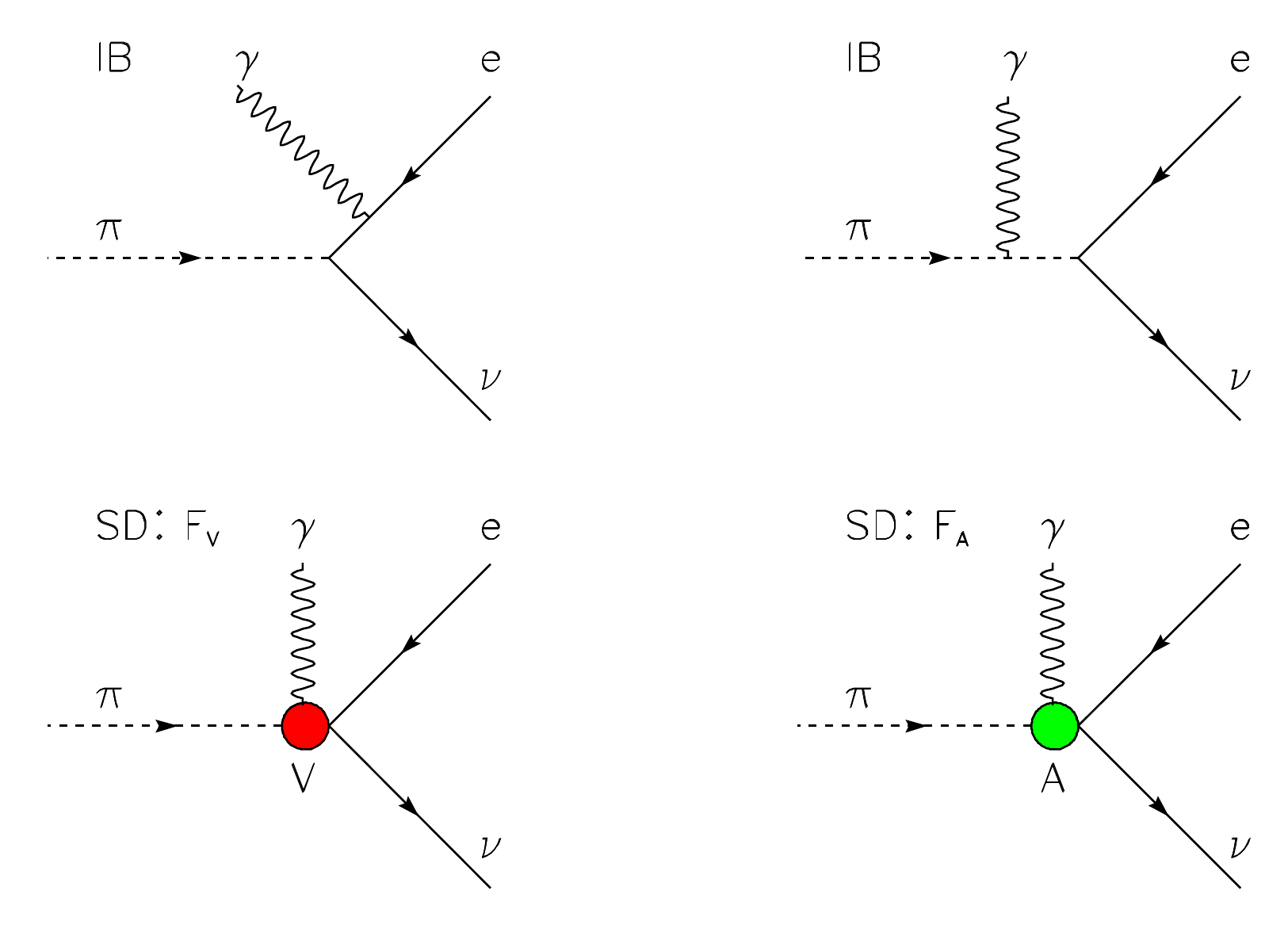}}
  \caption{Tree-level diagrams of the inner bremsstrahlung ($IB$) and
    structure-dependent ($SD$) amplitudes determined by the vector and
    axial-vector form factors, $F_V$ and $F_A$, respectively.  A new
    interaction type, such as one mediated by a hypothetical tensor
    particle proposed and discussed in the literature through the 1990s,
    would add an $SD$ amplitude defined by a corresponding form factor,
    i.e., $F_T$.}  \label{fig:pi_e2g_diags} 
\end{figure}

\begin{figure}[tb]
  \hspace*{0.13\linewidth}
    \parbox{0.4\linewidth}{\includegraphics[width=\linewidth]{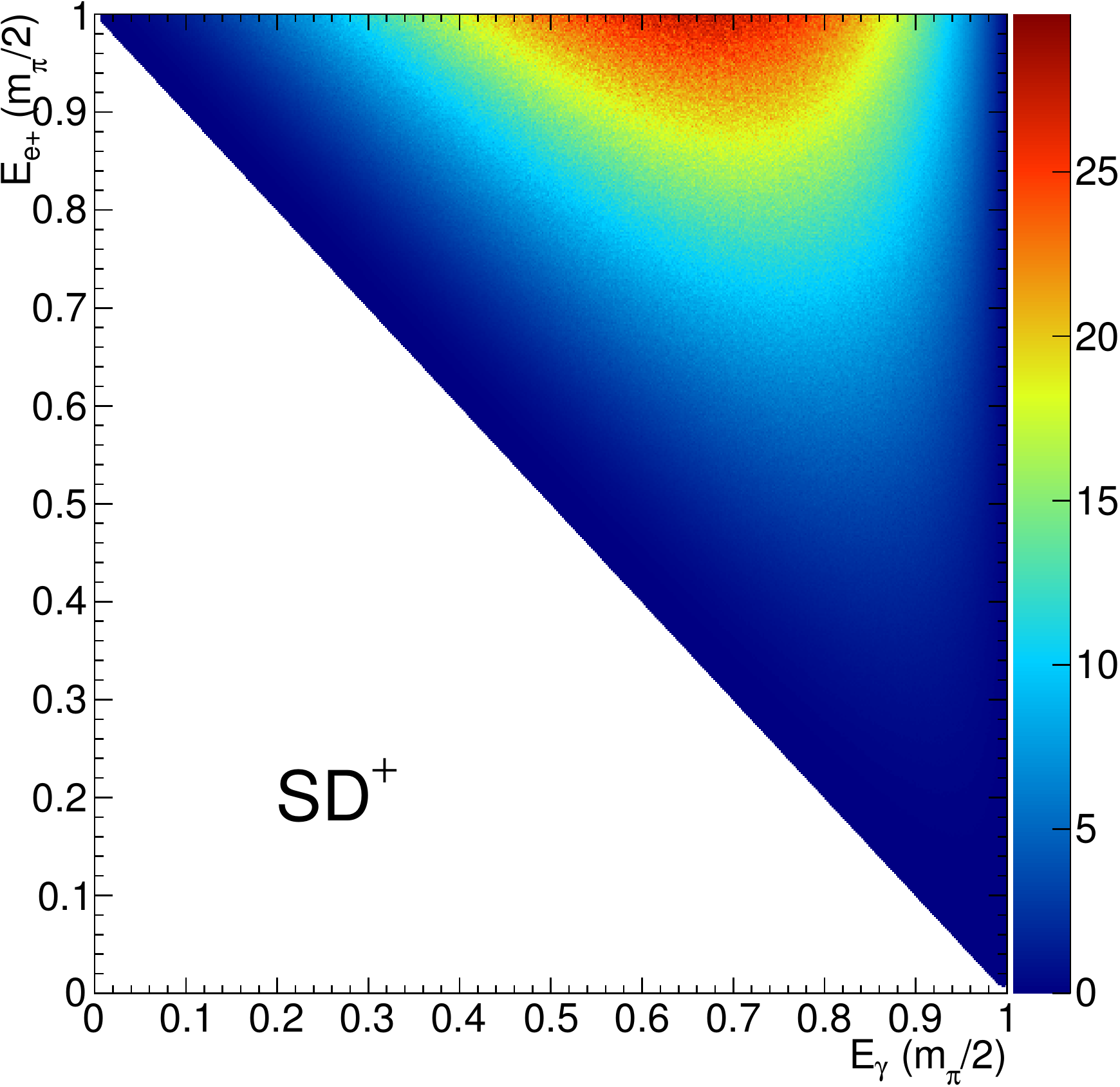}}
    \   
    \parbox{0.4\linewidth}{\includegraphics[width=\linewidth]{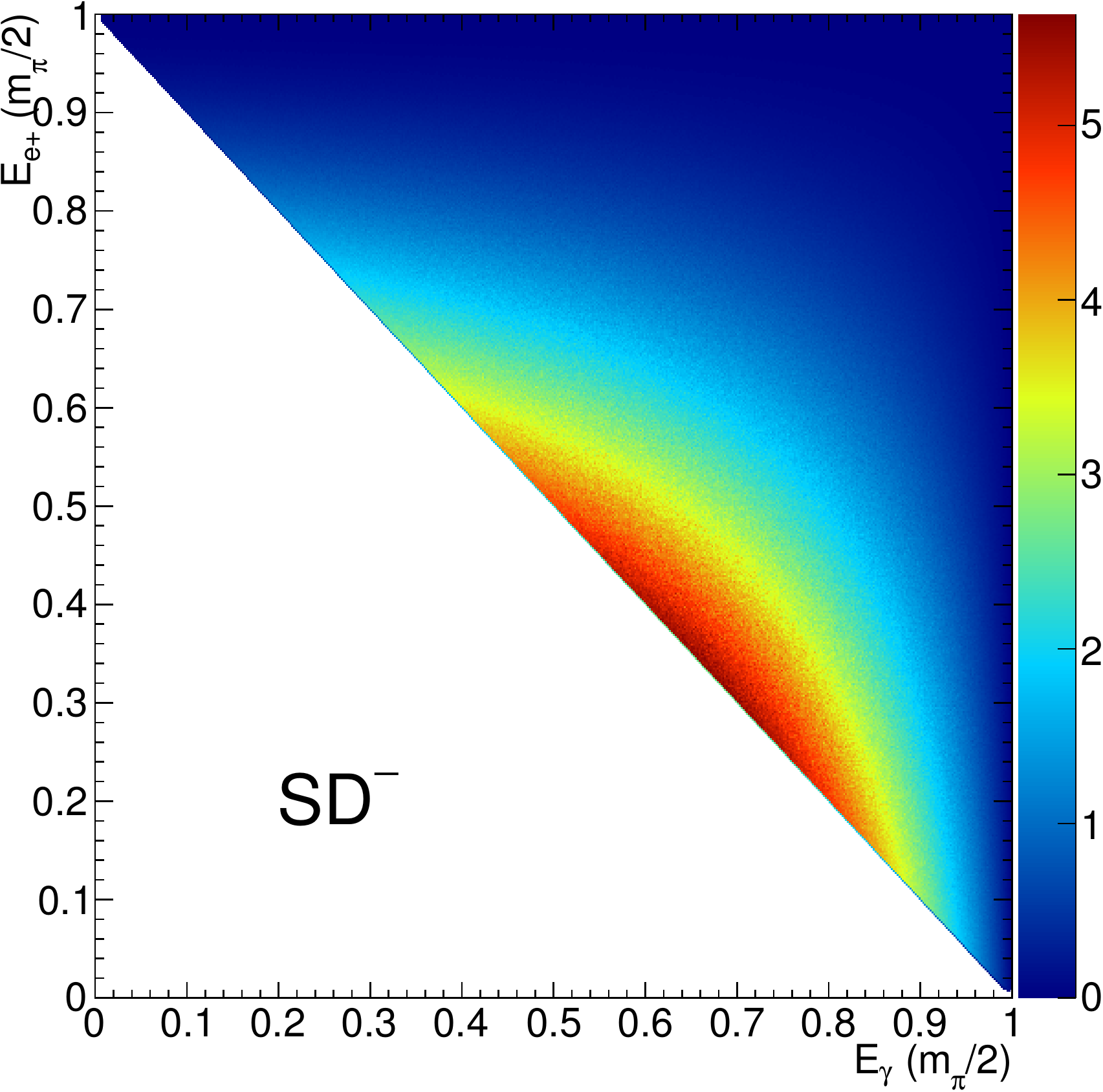}} \\
   \hspace*{0.13\linewidth}
    ~\parbox{0.41\linewidth}{\includegraphics[width=\linewidth]{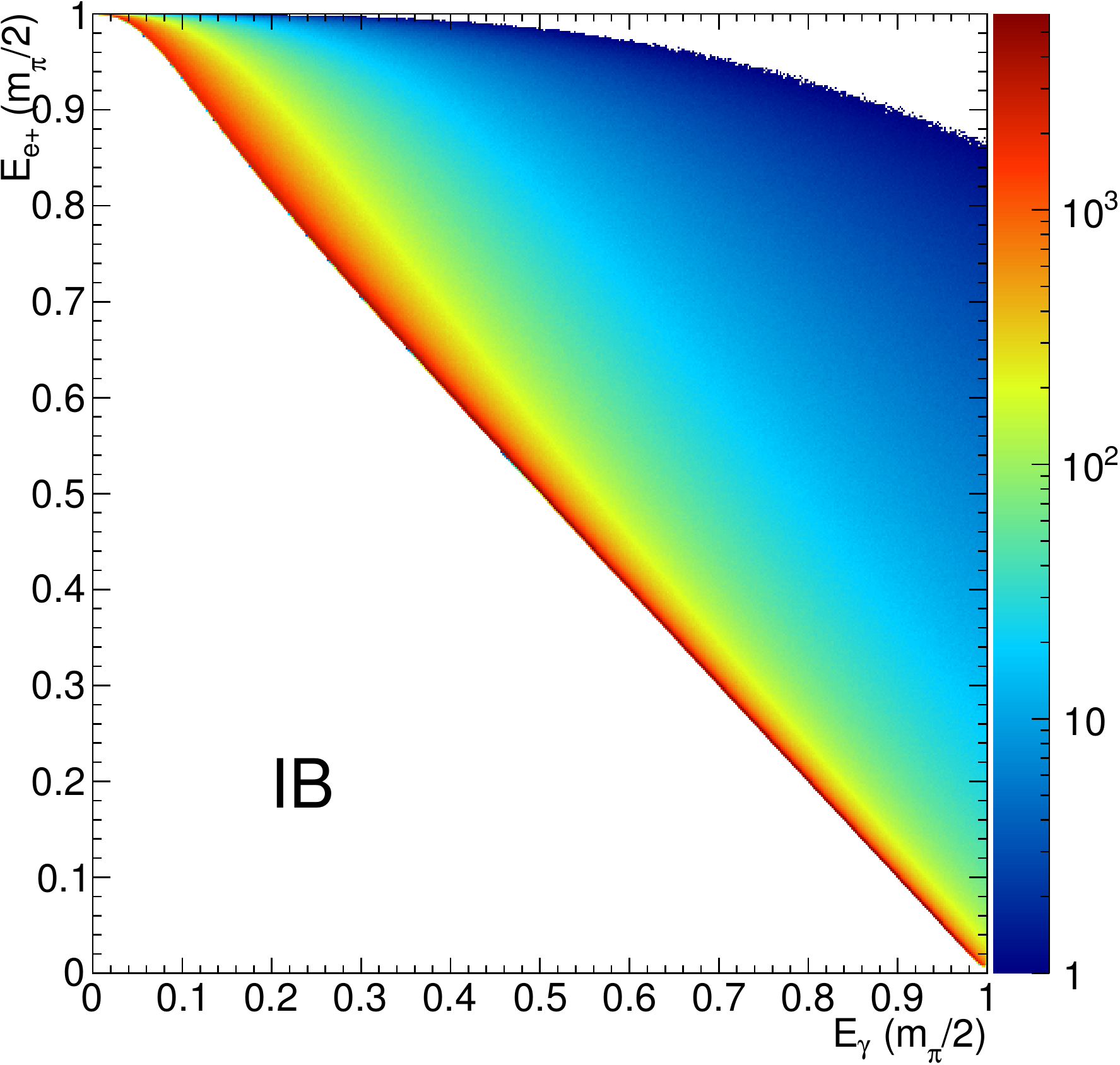}}
    \parbox{0.41\linewidth}{\includegraphics[width=\linewidth]{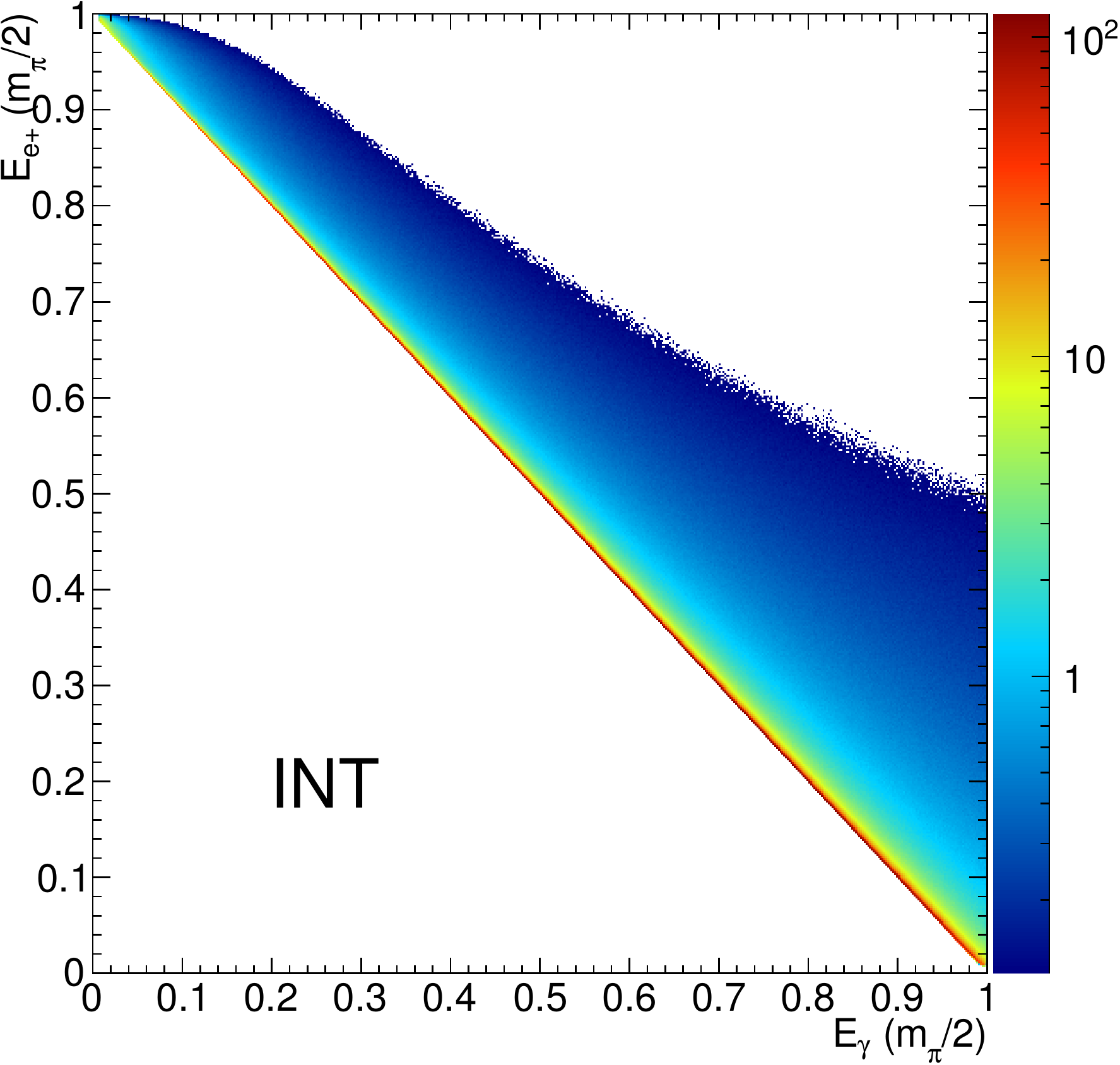}}
    \caption{Relative strengths of the physics amplitudes in $\pi\to
      e\nu\gamma$ decay: structure-dependent amplitudes $SD^+$ and
      $SD^-$, inner bremsstrahlung $IB$, and interference $INT \propto
      (1+\gamma) S_{\rm int}^+ + (1-\gamma) S_{\rm int}^-$, plotted as
      functions of $E_\gamma$ and $E_e$.  
          } 
  \label{fig:rpd_amplit}
\end{figure}

The decay $\pi^+ \to e^+\nu_e\gamma$ proceeds via a combination of QED
(inner bremsstrahlung, $IB$) and direct, structure-dependent ($SD$)
amplitudes \cite{Don92,Bry82}.  Under normal circumstances, as in
$\pi\to\mu\nu\gamma$ decay, the direct amplitudes are hopelessly buried
under an overwhelming inner bremsstrahlung background.  However, the
strong helicity suppression of the primary non-radiative process,
$\pi\to e\nu$, also suppresses the bremsstrahlung terms, making the
direct structure dependent amplitudes measurable in certain regions of
phase space \cite{Don92,Ber13}.  (We recall that the same helicity
suppression made possible sensitive searches for non-\VmA\ interaction
terms in precision measurements of the primary $\pi\to e\nu$ decay,
discussed in \sref{sec:pi_e2}.)  This result is of great use to
effective low-energy theories of the strong interaction, primarily ChPT,
which rely on the $SD$ amplitudes to provide important input parameters.
Whereas the $IB$ amplitude is completely described by QED, the
structure-dependent amplitude can be parametrized in terms of the pion
form factors.  As indicated in the tree-level Feynman diagrams
in \fref{fig:pi_e2g_diags}, standard \VmA\ electroweak theory requires
only two pion form factors, $F_A$, axial vector, and $F_V$, vector (or
polar-vector), to describe the $SD$ amplitude.
The amplitudes (form factors) $F_A$ and $F_V$ in principle depend
on the 4-momentum transfer to the $e$-$\nu$ pair (or to the $W$ boson),
$s=(p_e+p_\nu)^2\equiv q^2$.  In $\pi\to e\nu\gamma$ decay $s$ remains
low, $s<m_\pi^2$, so that it is a good approximation to evaluate $F_V$
and $F_A$ at $s=0$, often referred to as the soft pion limit.  It is
useful at this point to consider the tree-level double differential
branching ratio for the $\pi\to e\nu\gamma$ decay which, in the usual
parametrization, takes the form first worked out in detail by De~Baenst
and Pestieau \cite{DeB68}:
\begin{eqnarray}
  \fl
  \frac{\rmd^2\Gamma_{\pi e2\gamma}}{\rmd x\,\rmd y} = 
      \frac{\alpha}{2\pi} \Gamma_{\pi e2}
        \left\{ IB\left( x,y \right) 
       + \left( \frac{F_V m_\pi^2}{2 f_\pi m_e} \right)^{\!\!\!2}  
   \left[ \left(1 + \gamma \right)^2  SD^+(x,y)
      + \left( 1 - \gamma \right)^2  SD^-\left(x,y \right) \right]
        \right. \nonumber  \\
      + \left.\vphantom{\left(\frac{F_V m_\pi^2}{2 f_\pi m_e}\right)^2} 
      \left( \frac{F_V m_\pi}{f_\pi} \right) 
      \left[\left( 1+\gamma \right) S_{\rm int}^+\left( x,y \right) +
      \left( 1-\gamma \right) S_{\rm int}^- 
           \left( x,y \right) \right] \right\}\,,  
      \label{eq:RPD_SM_rate}
\end{eqnarray}
where $x\equiv 2E_\gamma/m_\pi$ and $y\equiv 2E_e/m_\pi$ are the
appropriately normalized photon and electron (positron) energies,
respectively, $\gamma\equiv F_A/F_V$ is the ratio of the axial and the
vector pion form factors, $f_\pi$ is the familiar pion decay constant
and ``int'' denotes interference terms between the $IB$ and $SD$
amplitudes.  The functional dependence of the terms $IB$, $SD^+$, $SD^-$
$SD^+_{\rm int}$ and $SD^-_{\rm int}$ on $x$ and $y$ is well established
and is given in the literature, e.g., in \cite{DeB68,Bry82}.  We note
that the $SD^+$ and $SD^-$ terms, which multiply the $(F_V+F_A)^2$ and
$(F_V-F_A)^2$ form factor terms, respectively, map very different
portions of the 3-body phase space in the final state.  The $SD^+$ term
peaks for large values of positron energies, $y \gtrsim 0.9$ and
moderate photon energies, $x \gtrsim 0.5$, where there is relatively
little background from the $IB$ terms.  The $SD^-$ term, on the other
hand, peaks for low values of $x+y \simeq 1$, where the $IB$ amplitude
is comparatively strong, as shown in \fref{fig:rpd_amplit}; this
kinematic region is also susceptible to background from muon radiative
decays.
It is not surprising, then, that most of the measurements of the
\peiig\ decay to date have focused on determining the $SD^+$ amplitude.

The conserved vector current (CVC) hypothesis~\cite{Fey58,Ger55} relates
the vector form factor $F_V$ to the $\pi^0\to\gamma\gamma$ decay
amplitude \cite{Vak58,Nev61,Mul63}:
\begin{equation}
  \Gamma_{\pi^0\to\gamma\gamma}
      = \frac{1}{2}\pi m_\pi\alpha^2\left|F_V\right|^2\,.
   \label{eq:F_V_from_CVC}
\end{equation}
During a long period of time ending in 2012, then current Particle Data
Group values for the $\pi^0$ lifetime and the $\pi^0\to\gamma\gamma$
branching fraction led to a CVC prediction of $F_V^{\rm CVC} =
0.0259 \pm 0.0009$.  As a rule, radiative pion decay measurements had
access only to one structure dependent amplitude, the $SD^+$.  Treating
the value of $F_V$ as known, predicted by the CVC hypothesis, allowed
analyses of \peiig\ data to extract $F_A$, or, more specifically the
ratio $\gamma\equiv F_A/F_V$.  Recent measurements of the $\pi^0$
lifetime, primarily that of the PrimEx collaboration \cite{Lar11}, have
led to a change in the PDG accepted value of the pion lifetime, along
with a reduced uncertainty.  Furthermore, Bernstein and Holstein have
recently pointed out the necessary correction to the CVC expression
in \eref{eq:F_V_from_CVC} arising from isospin breaking
effects \cite{Ber13}.  Combined, the two effects lead to a new isospin
corrected CVC prediction for the pion vector form factor:
\begin{equation}
   F_V^{\rm CVC} = 0.0255 \pm 0.0003\,,
   \label{eq:F_V-CVC_2013}
\end{equation}
which modifies the previously reported values for $F_A$ or $\gamma\equiv
F_A/F_V$.

Before discussing individual measurements of the radiative $\pi^+\to
e^+\nu_e\gamma$ decay, we note that the related decay $\pi^+\to
e^+\nu_ee^+e^-$, where the photon is virtual, adds $R$, a second axial
vector amplitude or form factor, to the $SD$ terms.  The $R$ form factor
is proportional to $\langle r_\pi^2\rangle$, the electromagnetic
(charge) radius of the pion squared \cite{Don92}.

\subsection{Measurements prior to the year 2000}

\subsubsection{\texorpdfstring {Ordinary radiative decay
    \peiig.}{Ordinary radiative decay pi (e2gamma).}}  The first
determination of the $SD$ amplitudes in \peiig\ decay was performed by
Depommier and collaborators.  In their apparatus pions were stopped in a
plastic scintillator target, and the decay positrons and photons
detected in opposing NaI(Tl) and lead glass detectors, each preceded by
a pair of thin plastic scintillator detectors in front \cite{Dep63b}.
Signals from all detectors were recorded on oscilloscope pictures which
were subsequently analyzed.  A total of $148 \pm 15$ \peiig\ events
resulted after background subtraction; some 110 were due to the $SD$
emission.  In their analysis, the authors used $F_V=0.0245$ and obtained
two solutions for $\gamma$, 0.4 and $-2.1$; the data did not favor one
over the other.  The authors also reported a branching ratio for
$E_e,E_\gamma >48$\,MeV with 17\% relative uncertainty, where, however,
the detector resolution had not been taken into account, so the result
could not be compared with theoretical calculations.

The Berkeley-UCLA experiment by Stetz and collaborators \cite{Ste78}
detected the positrons in a magnetic spectrometer and the photons in a
lead glass \v{C}erenkov hodoscope.  The experimenters observed $226 \pm
22.4$ events, and reported a branching ratio of $(5.6 \pm 0.7) \times
10^{-8}$ for $E_e>56$\,MeV and $\theta_{e\gamma}>132^\circ$.  To
determine the $SD$ terms, the authors used $F_V=0.0261 \pm 0.0009$, based
on the 1977 average value for the $\pi^0$ lifetime.  They, too, found
two solutions for $\gamma$: $0.44 \pm 0.12$ or $-2.36 \pm 0.12$,
neither of which was clearly favored by the data.

In the mid-1980's at the Swiss Institute for Nuclear Research (SIN, now
part of PSI) a Lausanne--Zurich collaboration \cite{Bay86} set up an
improved version of the Berkeley-UCLA apparatus to study the \peiig\
decay and determine $F_A$.  They used an intense pion beam with the
stopping rate in the target of $2.5 \times 10^7\,{\rm s}^{-1}$, an
8$\times$8 array of NaI(Tl) crystals for photon detection, and a
large-acceptance magnetic spectrometer for the positron.  An MWPC near
the target provided additional tracking information.  Data were taken in
two detector geometries, centered around $\theta_{e\gamma} = 180^\circ$
and $135^\circ$, respectively.  The former configuration accepts
negligible contributions from the $IB$ and $SD^-$ amplitudes, while the
latter accepts a sizable $SD^-$ contribution still with minimal $IB$
background.  Using this method, the collaborators collected $653 \pm
29$ \peiig\ events, as well as $801 \pm 34$ \peii\ events for
normalization.  Thanks to the expanded phase space coverage compared to
previous experiments, Bay \etal\ were able to resolve the quadratic
ambiguity inherent in the $SD^+$ analysis only, and eliminate the
negative-sign solution by a confidence factor of 8.5.  Using the CVC
value of $F_A=0.0255(5)$, incidentally the same central value as the
current best isospin-corrected CVC prediction in \eref{eq:F_V-CVC_2013},
the authors obtained $\gamma=0.52 \pm 0.06$, or $F_A=0.0133 \pm 0.0015$.

Around the same time as the SIN experiment, a group at the Los Alamos
Meson Physics Facility (LAMPF) made a measurement of $F_A/F_V$
\cite{Pii86} using the Crystal Box detector which consisted of 396
NaI(Tl) crystals, 36 plastic scintillation hodoscope counters, and a
cylindrical drift chamber surrounding the stopping target \cite{Bol84}.
Unlike the other experiments discussed above, the Crystal Box detector's
single-particle acceptance for events originating in the target was
about 45\% of $4\pi$\,sr.  Such a large acceptance enabled the
experimenters to cover a broad portion of the decay phase space,
collecting 2364 coincidence events that included both decay signals and
random coincidences.  However, only $71 \pm 11$ high-energy nearly
back-to-back photon-positron pairs were used to determine $\gamma$, as
they were largely background free and strongly dominated by the $SD^+$
amplitude.  The low statistics is reflected in the large relative
uncertainty in $\gamma= 0.25 \pm 0.12$; however the remaining $\sim$2300
coincident events were included in a maximum likelihood analysis which
preferred the positive $\gamma$ solution by a factor of 2175:1.  In all,
the likelihood analysis found a total of $179 \pm 18$ $\pi\to
e\nu\gamma$ events, in good agreement with the integral of the
$e$-$\gamma$ timing peak.  Thus, although a low statistics measurement,
this experiment's main result is the strongly favored selection of the
$\gamma>0$ solution, removing the prior longstanding quadratic
ambiguity.

Although the subject of this review are measurements of rare decays of
the pion, it is worthwhile to consider the 1988 analysis of Dominguez
and Sol\`a \cite{Dom88} who extracted a soft-pion value for $F_A(0)$
from semileptonic tau lepton decays, $\tau\to\nu_\tau+n\pi$.  Analysis
of decays with odd (even) values of $n$ gives access to the vector
(axial-vector) hadronic spectral functions up to the kinematical limit
$t \sim 3\,{\rm GeV}^2$.  Dominguez and Sol\`a studied and refined such
fits to existing $\tau$ decay data, and extracted the soft-pion value of
$F_A(0)=0.017 \pm 0.004$, or $\gamma(0) = F_A(0)/F_V(0) = 0.67 \pm
0.17$, with $F_V^{\rm CVC} \simeq 0.0254$.  We note that this result
is \emph{independent of pion radiative decay} data.

The lone measurement of $\pi\to e\nu\gamma$ decays in flight was
performed by the Moscow Institute for Nuclear Research (INR) group,
using the ISTRA apparatus and a 17\,GeV pion beam at the IHEP Protvino
U70 accelerator \cite{Bol90}.  The experimental technique and
systematics are radically different from all of the stopped pion decay
measurements discussed so far.  Since decays occurred in flight, the
experimenters were free to use negative pions and observe $\pi^-\to
e^-\nu\gamma$ decays, making this the only experiment to date to do so.
The wide acceptance of the apparatus enabled the experimenters to study
the pion rest frame kinematical region defined by
\begin{equation}
   E_\gamma>21\,{\rm MeV},\quad
   E_e> 70\,{\rm MeV} -0.8E_\gamma, \quad {\rm and} \quad
   \theta_{e\gamma}>60^\circ\,.  \label{eq:Bol_kinematics}
\end{equation}
The results of this work, obtained in a maximum likelihood analysis, can
be summarized as follows: (a) $\gamma\equiv F_A/F_V=0.41 \pm 0.23$ (with
no corresponding value for $F_V$ quoted explicitly), (b) negative
solution for $\gamma$ disfavored by a likelihood factor of $5\times
10^9$ or 6.7 standard deviations, (c) an independently determined
$F_V=0.014 \pm 0.009$, and (d) the branching ratio for the kinematic
limits of equation \eref{eq:Bol_kinematics}, $B= (1.61 \pm 0.23)\times
10^{-7}$.  Although not explicitly given, the total number of $\pi\to
e\nu\gamma$ decay events appears to be approximately 90 after background
subtraction, which would explain the large quoted uncertainties.
Besides the strong preference for the $\gamma>0$ solution, this work
presented another notable result: a deficit of $SD^-$ events, forcing a
non-physical negative $SD^-$ amplitude.  The authors speculated that the
deficit may be due to a destructive interference with a tensor term of
the size $F_T = 0.0056 \pm 0.0017$.  We will revisit the tensor
hypothesis below.

\subsubsection{\texorpdfstring
      {The decay $\pi^+\to e^+\nu_ee^+e^-$ (or \peiiee)} {The decay pi+
        to e+ nu e+ e- [or pi(e2ee)]} } 
merits special mention, as it contributes information not accessible
through the regular \peiig\ radiative decay channel.  Following
unsuccessful attempts \cite{Kor76}, the \peiiee\ decay was first
observed by the SINDRUM I collaboration in a 1985 experiment
\cite{Egl86}; the same data were reanalyzed more carefully and the
results reported in 1989 \cite{Egl89}.

\begin{figure}
   \includegraphics[width=\linewidth]{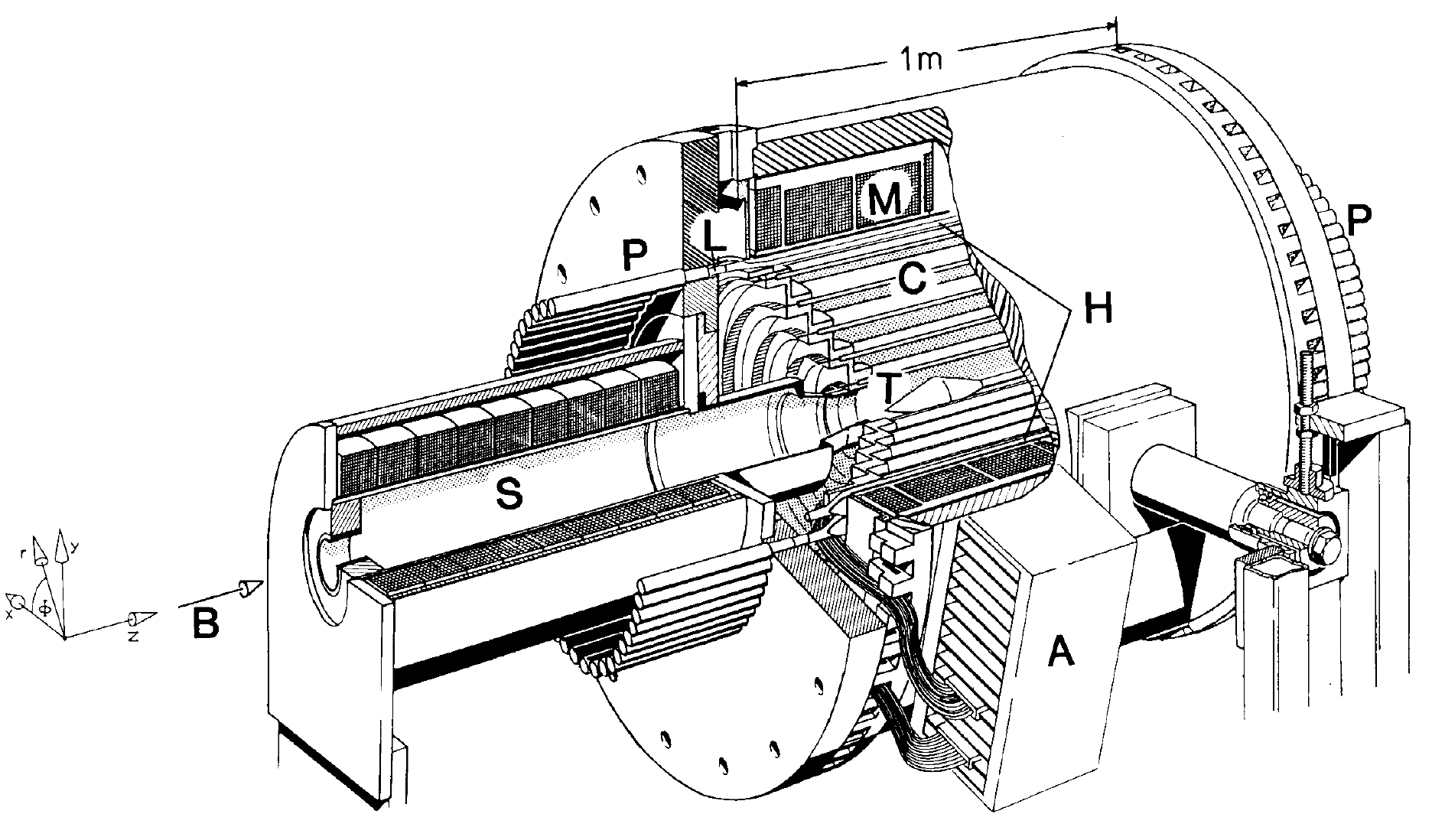}
   \caption{Schematic view of the SINDRUM I spectrometer as originally
     configured for the $\mu\to eee$ search \cite{Ber85}: B, $\mu$ beam;
     S, focusing solenoid (not used with the pion beam); T, target (a
     different target was used for the \peiiee\ measurement); C, five
     cylindrical MWPC's; H, hodoscope of 64 scintillators; L, light
     guides for the hodoscope; P, 128 photomultiplier tubes; A, cathode
     strip preamplifiers and anode wire amplifier/discriminators; M,
     magnet coil.  For further details see \cite{Ber85,Egl86,Egl89}}
   \label{fig:sindrum}
\end{figure}

The SINDRUM detector system, which became known as ``SINDRUM I'' after
the construction of ``SINDRUM II'', was built primarily to search for
the forbidden $\mu\to 3e$ decay \cite{Ber85}.  The instrument,
schematically depicted in \fref{fig:sindrum}, was capable of recording
the trajectories of electrons and positrons in a solenoidal magnetic
field with the help of five concentric cylindrical MWPC's of very
similar design and construction to those used later in the PIBETA/PEN
experiments.  Decay particles were required to reach a 64 element
scintillating hodoscope situated outside the tracking detectors,
resulting in a lower threshold on transverse momentum around
17\,MeV/$c$.  A cone-shaped 12-element segmented scintillating target
was used, enabling a reduction of random coincidences between beam pions
and decay particles by an order of magnitude.

In their first paper Egli \etal \cite{Egl86} reported merely the first
observation of the decay, with 79 recorded events.  The authors also
reported values of form factor ratios $\gamma=F_A/F_V$ and $R/F_V$ with
large uncertainty limits.  They were, however, able to exclude by a
ratio of 9:1 the previously reported large negative value of $\gamma =
-2.49\pm0.06$ from \peiig\ \cite{Ste78} in favor of a smaller positive
solution $\gamma = 0.7\pm0.5$, thus anticipating the soon to follow
\peiig\ results of the Lausanne--Zurich \cite{Bay86} and LAMPF
\cite{Pii86} groups.

In a second article \cite{Egl89}, the SINDRUM I collaboration reported
results of a more careful and comprehensive analysis of the same 1985
data set.  By applying a series of cuts designed to reduce the
backgrounds, the authors arrived at a data set of 98 events with 1
background event, based on $\sim 4 \times 10^{12}$ pions stopped in the
target.  Using this event set, the detector acceptance, the CVC value
$F_V=0.0255$, the Bay \etal value $F_A=0.0133 \pm 0.015$ \cite{Bay86}
and the PCAC value for the second axial form factor $R_{\rm PCAC}=0.068
\pm 0.004$, the authors derive the branching ratio for the full phase
space:
\begin{equation}
   B(\pi^+\to e^+\nu_e e^+e^-)= (3.2\pm 0.5 \pm 0.2)\times 10^{-9}\,,
\end{equation}
where the second uncertainty is propagated from $R_{\rm PCAC}$.  Of
additional interest are the values for the three pion form factors
derived in a likelihood analysis of the data:
\begin{equation}
   F_V = 0.023 ^{+0.015}_{-0.013}\,, \qquad
   F_A = 0.021 ^{+0.011}_{-0.013}\,, \qquad
   R = 0.059 ^{+0.009}_{-0.008}\,.
\end{equation}
It is worth noting that fixing the vector form factor value to $F_V^{\rm
  CVC}=0.0255$ does not significantly alter the maximum likelihood
values of $F_A$ and $R$.

A subsequent Dubna experiment \cite{Bar92} which detected only 7 events
of \peiiee\ decay did not improve the precision of the SINDRUM I
branching ratio for the decay.

\subsubsection{Summary of the early measurements:} The $\pi^+\to
e^+\nu_e\gamma$ and $\pi^+\to e^+\nu_ee^+e^-$ results published  prior
to 2000 may be summarized as follows. 
\begin{enumerate}
  \renewcommand{\labelenumi}{(\roman{enumi})}

  \item The positive-sign solution for $\gamma\equiv F_A/F_V$ was
    established with high likelihood by three experiments, effectively
    ruling out a negative solution.

  \item The number of reconstructed events adds up to less than 1,200
    events~\cite{Dep63b,Ste78,Bay86,Pii86,Bol90}.  The combined
    statistical and systematic uncertainties of the parameter
    $\gamma \equiv F_A/F_V$ extracted by the individual experiments
    range from 12\,\%~\cite{Bay86} to 56\,\%~\cite{Bol90}.
    The world average \cite{PDG04}
    \begin{equation}
       F_A^{\rm pre\mbox{-}2000} = 0.0116 \pm 0.0016 \,,
             \label{eq:F-A_pre-2000}
    \end{equation} was assigned the confidence level of 0.175 by the
    PDG. 

  \item There were only two, relatively low precision measurements of
    $F_V$ resulting in a world average with a 50\% uncertainty:
    \begin{equation}
       F_V^{\rm pre\mbox{-}2000} = 0.017 \pm 0.008 \,,
       \label{eq:F-V_pre-2000}
    \end{equation}

  \item There was only one, low precision measurement of the $\pi\to
    e\nu\gamma$ branching fraction, defined over a correlated kinematic
    region given in equation \eref{eq:Bol_kinematics}.

  \item The Protvino experiment \cite{Bol90} raised the possibility of a
    substantial tensor term, $F_T$, subsequently discussed during more
    than a decade in a series of theoretical papers \cite{Chi05}.

\end{enumerate}

\subsection{\texorpdfstring
   {PIBETA measurements of the $\pi^+\to e^+\nu_e\gamma$ decay}
   {PIBETA measurements of the pi+ to e+ nu(e) gamma decay}
   \label{sec:pibeta-rpd}}

Study of the radiative pion decay has been a major component in the long
term program of the PIBETA/PEN experiments at PSI.  Firstly, the strong
intrinsic physics significance for low-energy QCD and ChPT as well as
the sensitivity to non-\VmA\ interaction terms such as tensor, and its
relatively poor experimental quantification prior to 2000 discussed in
the preceding section, clearly placed radiative pion decay at high
priority.  Secondly, $\pi^+\to e^+\nu\gamma$ events for which the
positron annihilates externally create a significant background in the
measurement of the pion beta decay $\pi^+\to\pi^0e^+\nu$, \peiii, the
primary goal of the PIBETA experiment, discussed in \sref{sec:pi-beta}
below.  Thirdly, \peiii\ events for which one of the photons in the
subsequent $\pi^0$ decay converts into an asymmetric $e^+e^-$ pair, with
only one detected electromagnetic shower, create a background for the
\peiig\ process.  Furthermore, a precise measurement of $R_{e/\mu}^\pi$
of equation \eref{eq:pi_e2_general} is impossible without a precise
knowledge of the radiative decay width.  In fact, because of the
overlapping nature of the instrumental response functions to the various
decay processes, precision measurement of any one of them requires
simultaneous detection and characterization of all pion and muon decay
processes present in the data sample.

\begin{figure}[htb]
   \centerline{ \includegraphics[width=0.4\linewidth]{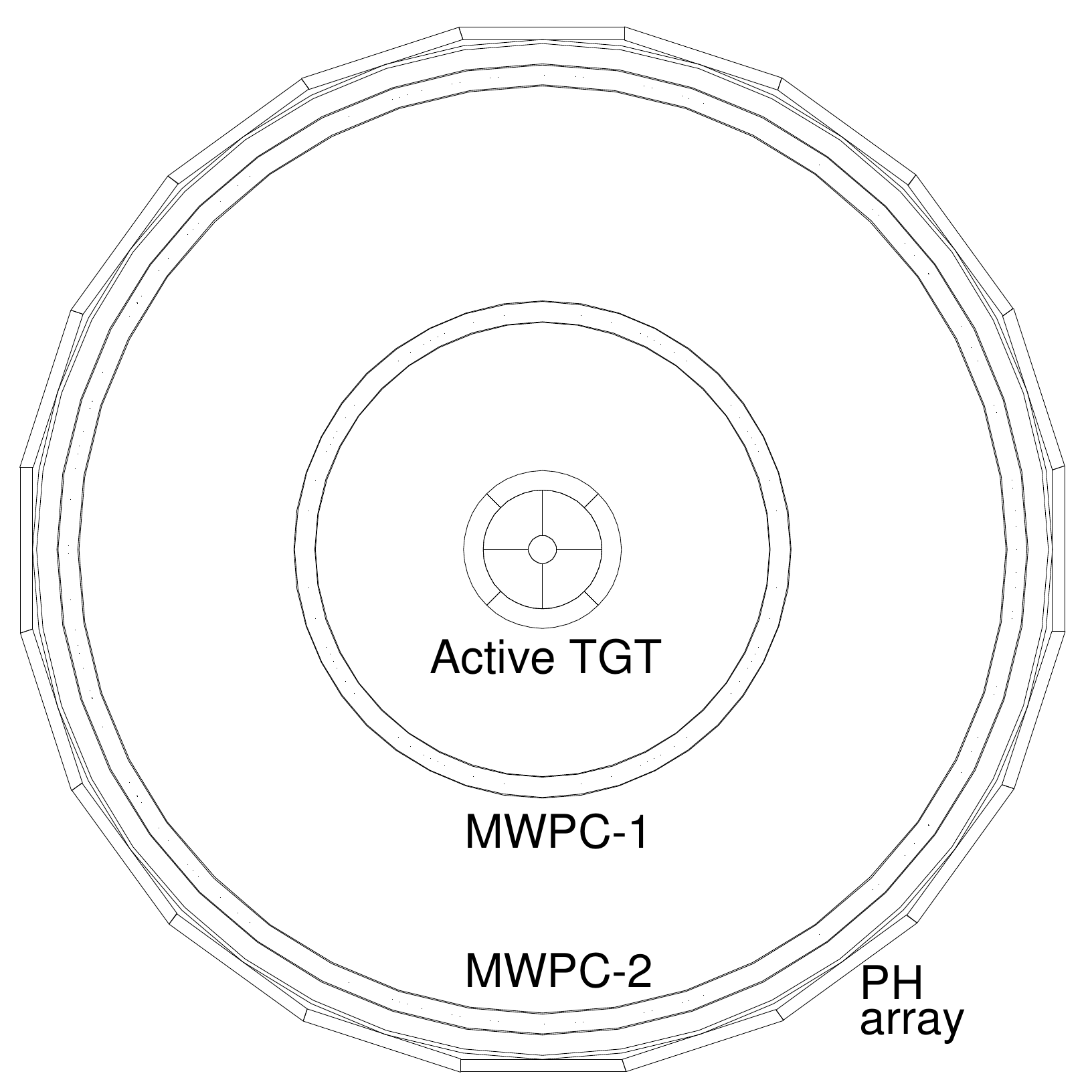} }
   \caption{Axial (beam) view of the central detector region during the
     first PIBETA running period, 1999-2001, and the first half of the
     2004 run.  Going outward from center we first note the 9-element
     active target (AT), followed by multiwire proportional chambers
     MWPC1 and MWPC2, a thin carbon-fiber mechanical shield, and,
     finally, the 20-element plastic hodoscope (PH).  The circumscribed
     radius is approximately 15\,cm.}
     \label{fig:cent_det}
\end{figure}

For all these reasons, the PIBETA collaboration \cite{PBweb} performed a
series of measurements at the Paul Scherrer Institute, focused on
improving the experimental precision of the \peiig\ branching ratio as
well as the form factors $F_A$ and $F_V$.  The apparatus is essentially
the same as in the PEN configuration shown in \fref{fig:PEN_det}.  The
only differences were the absence of the beam mini time projection
chamber (mTPC), the use of a segmented target in the early runs
(\sref{sec:pi-beta}), a thicker active target and degrader detectors to
accommodate the higher momentum of $\sim$114\,MeV/$c$, pion beam, and
custom electronics for 1- and 2-arm triggers.  The arrangement of the
central detectors is shown in \fref{fig:cent_det}.  The data were
collected in two distinct sets of runs.

\begin{figure}[htb]
   \parbox{0.50\linewidth}{
           \includegraphics[width=\linewidth]{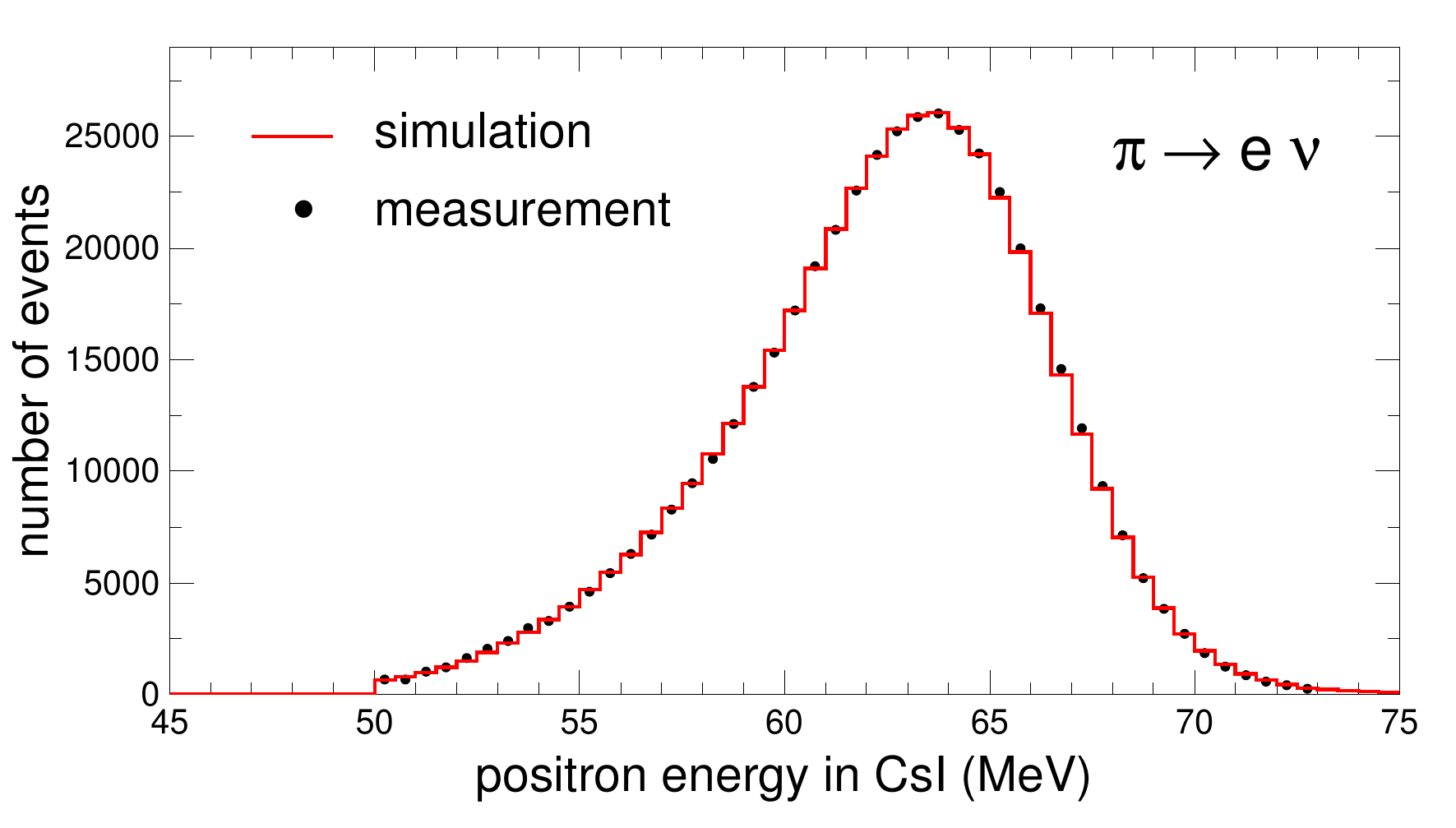}}
   \hspace*{\fill}
   \parbox{0.47\linewidth}{
            \includegraphics[width=\linewidth]{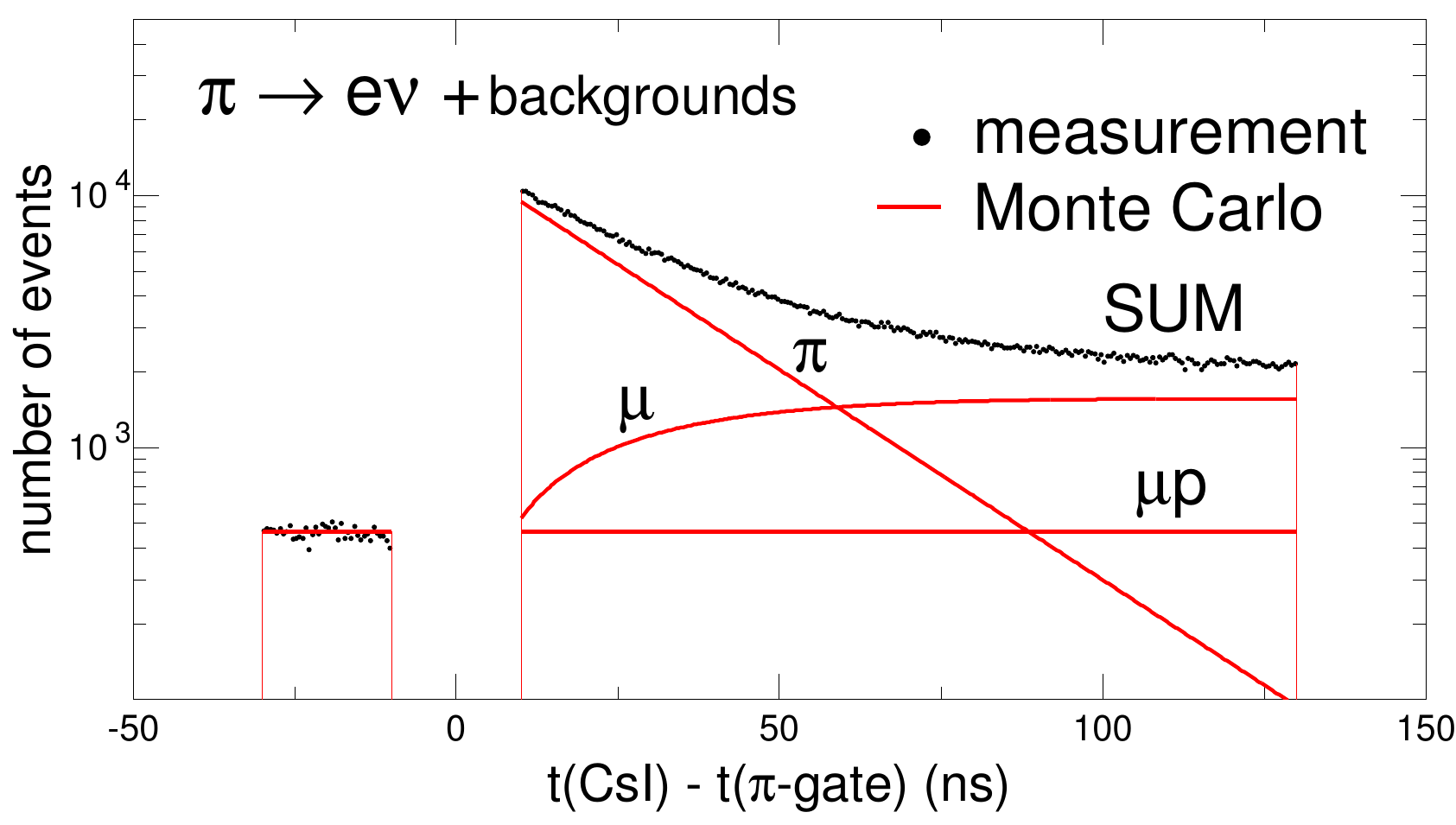}}
   \caption{CsI calorimeter response for $\pi\to e\nu$ decays used for
     calibration and normalization in the PIBETA experiment.  Left panel
     shows the net, background-subtracted energy response for
     high-threshold events.  Energies below $\sim 50$\,MeV were
     dominated by the Michel $\mu^+\to e^+\nu\bar{\nu}$ decays, and are
     not shown.  The red histogram depicts results of a realistic GEANT3
     Monte Carlo simulation.  Right panel shows the time response for
     single-arm high threshold events including backgrounds.  Time $t=0$
     is determined by the pion stop; data taking was stopped for
     approximately 10\,ns around the pion stop time (events during this
     ``{\sc prompt}'' gate were collected with a high prescaling
     factor).  The main backgrounds come from muon decay, either causal
     ($\mu$) through the $\pi\to\mu\to e$ decay chain, or acausal
     ($\mu$p) from muons piled-up in the target region.  The latter were
     determined precisely through event triggers for which the decay
     positron preceded the pion stop signal by up to $\sim
     30$\,ns.}  \label{fig:PB_e2_E-t_resp}
\end{figure}

The first PIBETA radiative pion decay measuring period took place during
1999--2001, in a configuration optimized for recording pion beta decay,
$\pi^+\to\pi^0e^+\nu_e$, events with a high stopping rate of pions in
the target, averaging approximately $8\times 10^5\,{\rm s}^{-1}$.  The
experiment trigger logic was based on shower energy summed over a
cluster of adjacent CsI calorimeter detector modules, with two energy
thresholds: a high one of $\sim 51$\,MeV, and a low one of $\sim
5$\,MeV.  The experiment recorded every event containing two showers,
both with energy above the high threshold (HT), and spatially separated
by an opening angle $\theta_{12}\gtrsim 90^\circ$.  Two-shower events
with one exceeding the high and the other the low energy threshold were
prescaled by a factor of at least an order of magnitude, while events
with two low-threshold showers were prescaled even more.  One-arm
events, consisting of one shower were also recorded, with the purpose of
collecting significant samples of $\pi^+\to e^+\nu$ and $\mu^+\to
e^+ \nu\bar{\nu}$ events for normalization and systematic studies.
Detector response to the $\pi\to e \nu$ single-arm events, shown
in \fref{fig:PB_e2_E-t_resp}, best illustrates the intrinsic performance
of the spectrometer.
Over $4 \times 10^4$ \peiig\ events were collected and analyzed in this
set of runs.  The results, bringing about an order of magnitude
improvement in precision of the branching ratio and $F_A/F_V$, were
published in \cite{Frl04b}.  However, the authors found a deficit of
events in one kinematic region, corresponding to high-$E_\gamma$ and
low-$E_e$ events, that was well outside the limits of the statistical
uncertainties of the fit.  The authors concluded that further study was
needed in a dedicated measurement, as the first run had been optimized
for measurement of $\pi^+\to\pi^0e^+\nu_e$ decays.

\begin{figure}[htb]
  \centerline{\includegraphics[width=0.8\linewidth]{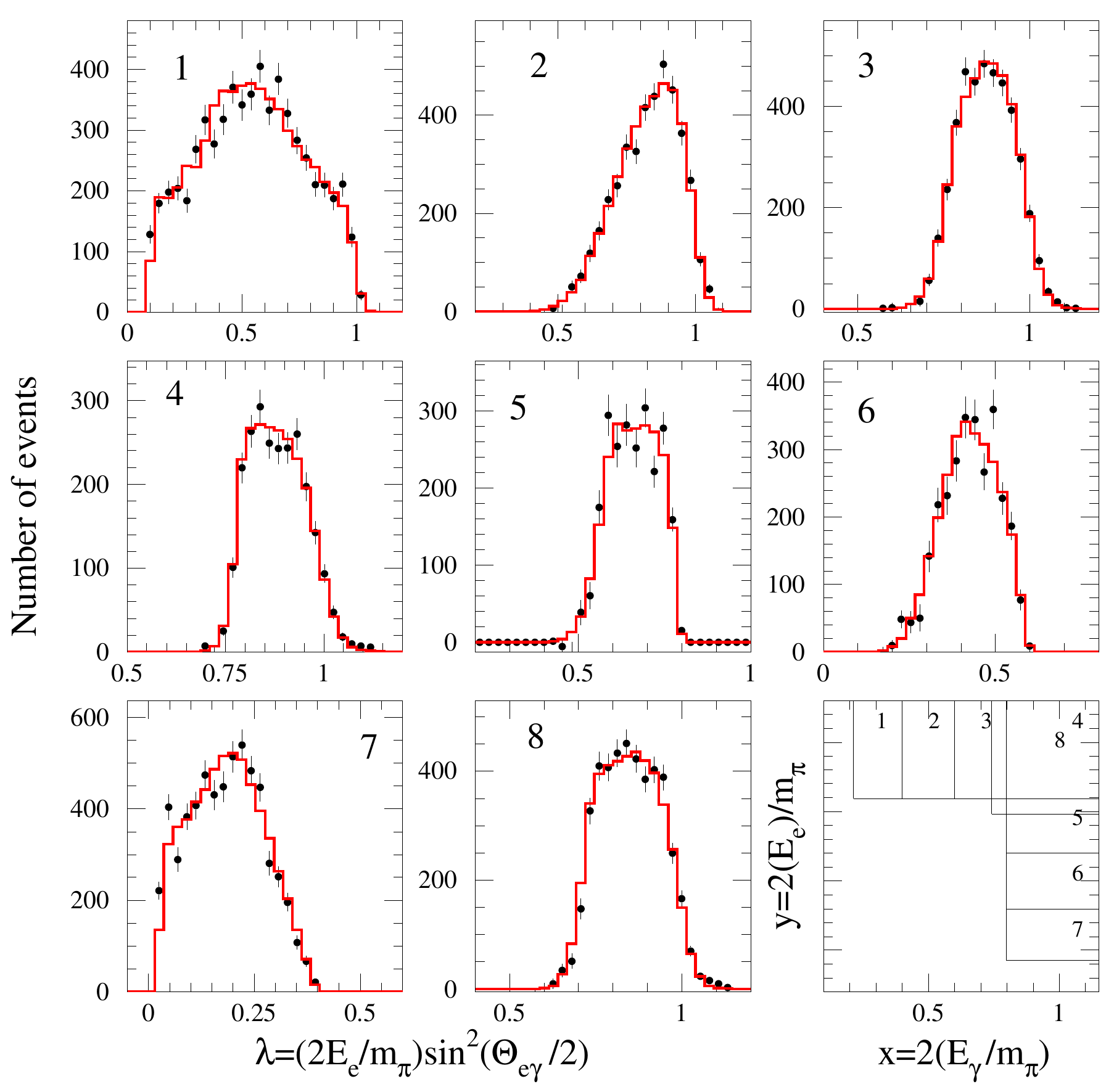}}
  \caption{PIBETA data points: background-subtracted $\pi^+\to
    e^+\nu_e\gamma$ distribution of the kinematic variable
    $\lambda \equiv \left(2E_e/m_\pi \right) 
      \sin^2\left( \theta_{e\gamma}/2 \right)$ for 8 regions in the
    $E_e$-$E_\gamma$ (or $y$-$x$) plane.  The layout of the regions in
    the $x$-$y$ plane is indicated in the lower right panel.  Solid red 
    histograms: results of GEANT3 calculations for the best-fit values
    of $F_V$, $F_A$, and $a$.  } \label{fig:pienug_regions}
\end{figure}

\begin{figure}[htb]
  \centerline{\includegraphics[width=0.7\linewidth]{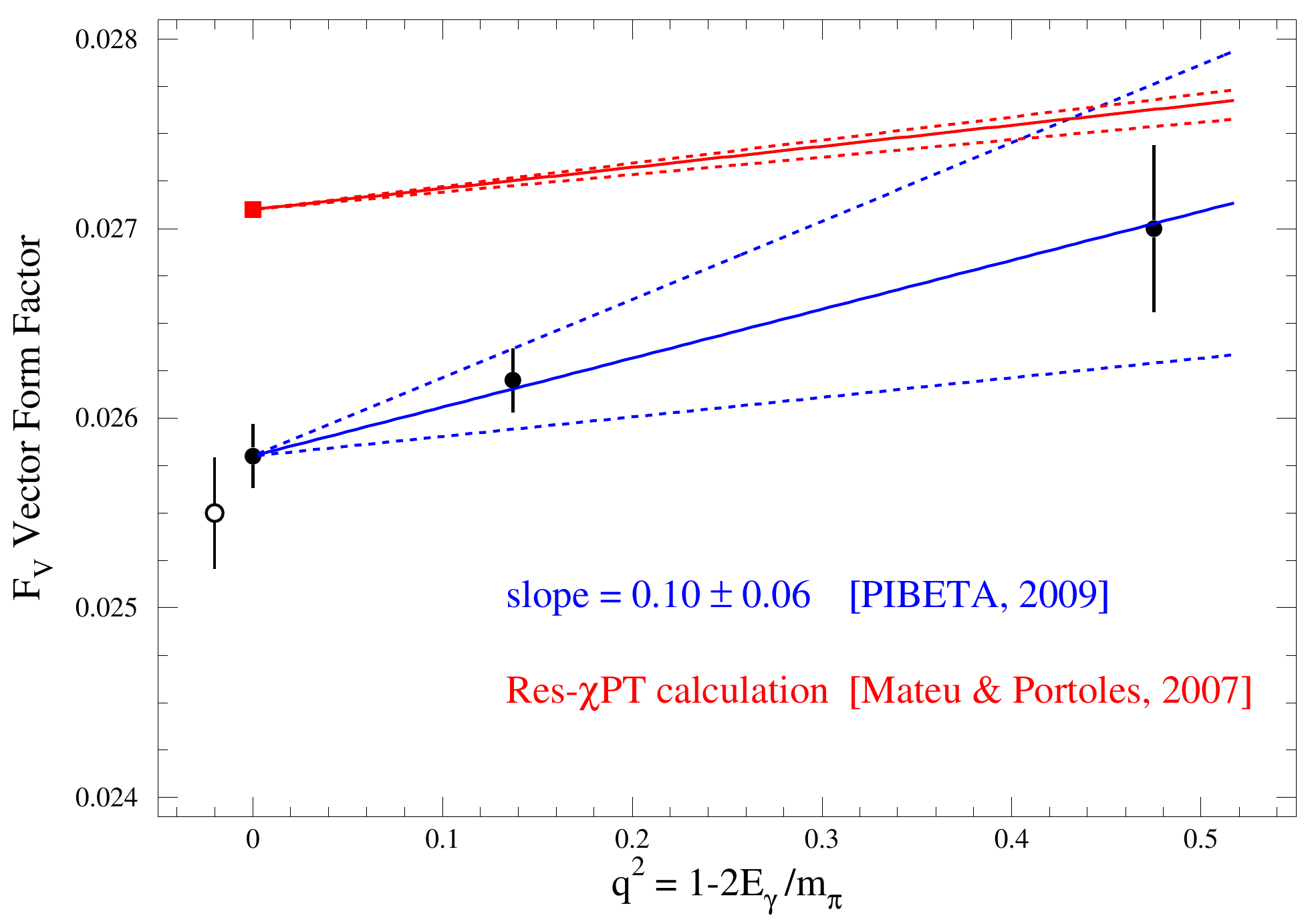} }
  \caption{Details of the $q^2=(p_e+p_\nu)^2$ slope analysis for
    $F_V(q^2)$ in the PIBETA \peiig\ measurement \cite{Byc09}.  Black
    data points: best-fit values of $F_V(q^2)$ for three momentum bins
    in the measurement.  Solid and dashed blue lines represent the
    central values and $1\sigma$ band, respectively, of the minimum
    $\chi^2$ fit.  Solid red square point and red lines: results of the
    resonance ChPT calculation of Mateu and Portol\'es \cite{Mat07}.
    The current isospin-corrected CVC prediction $F_V(0) = 0.0255 \pm
    0.0003$ is indicated by the empty circle offset slightly to $q^2<0$
    for better visibility.}
  \label{fig:slope-2013}
\end{figure}

\begin{figure}[htb]
  \centerline{\includegraphics[width=0.7\linewidth]{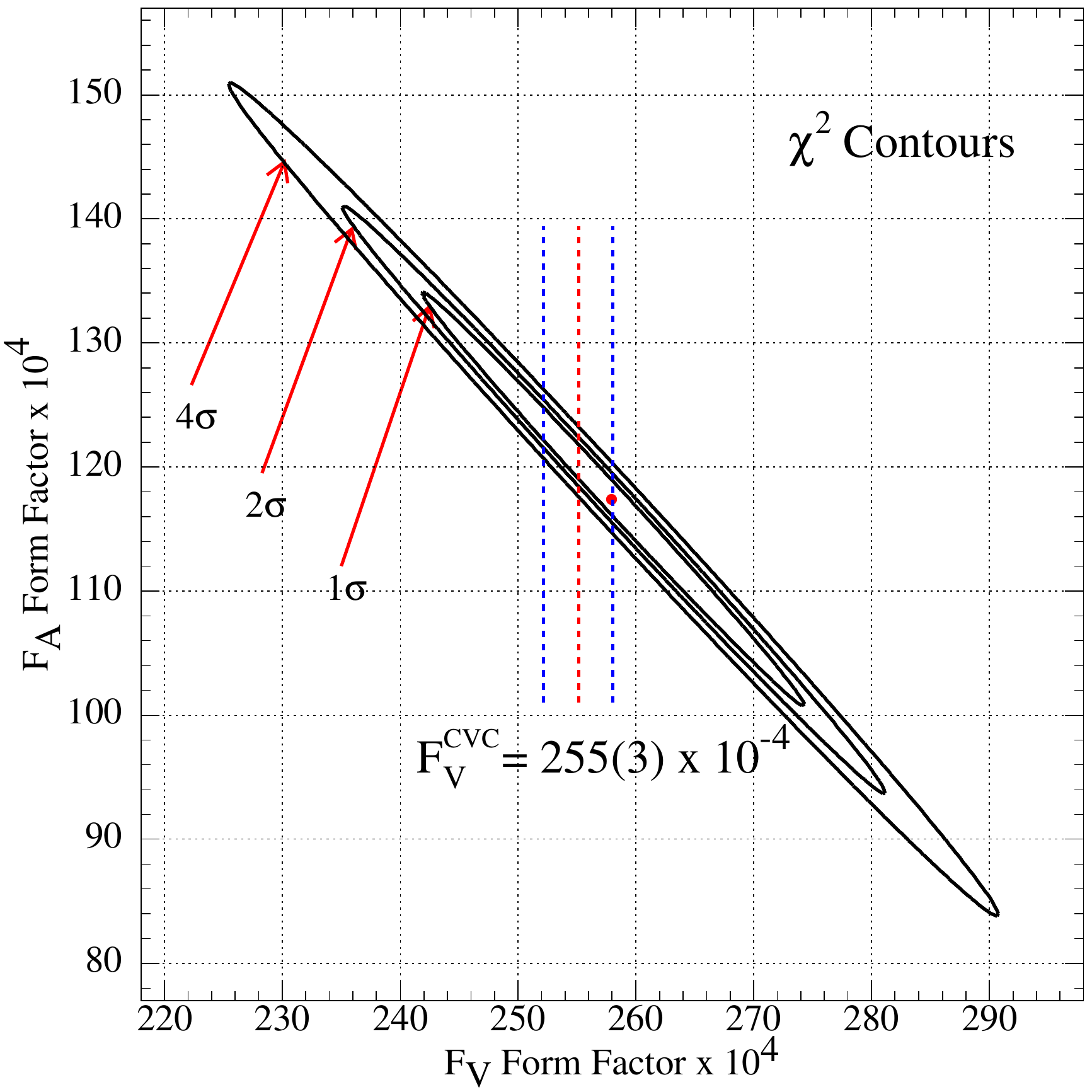} }
  \caption{Contour plot of loci of constant $\chi^2$ for the minimum
    value $\chi^2_0$ (solid red dot) plus 1, 2, and 4 units,
    respectively, in the $F_A - F_V$ parameter plane, keeping the slope
    parameter $a$ fixed at the theoretical value $a = 0.041$
    \cite{Mat07}.  The range of the CVC prediction $F_V = 0.0255 \pm
    0.0003$ is indicated by the dashed vertical lines.}
  \label{fig:F_A-F_V-2013}
\end{figure}

The second run, dedicated to radiative \peiig\ decays took place in
2004, with the pion stopping rate lowered to $1.5\times 10^5\,{\rm
s}^{-1}$ and the trigger electronics modified to accept all
high-threshold one-arm events.  The first half of the run was carried
out with the same detectors as in 1999-2001, while in the second half
the 9-element active target was replaced with a single unit.
Approximately $2.4 \times 10^4$ additional events were collected over
significantly broader kinematic regions.  Most importantly, the lower
beam rate resulted in large improvements in the ratio of peak signal to
accidental background (P/B) in $\Delta t_{e\gamma}$ spectra, and
sufficient low-energy data to perform an independent energy calibration
for the charged and neutral showers in the HT trigger.  The combined
data sets were carefully analyzed, and the results published in
\cite{Byc09}.  The key results are shown in figures
\ref{fig:pienug_regions}, \ref{fig:slope-2013}
and \ref{fig:F_A-F_V-2013}. 

With the improved calibrations, the previously observed event deficit
vanished, as seen in \fref{fig:pienug_regions}, which shows the measured
yield in 8 different kinematic regions, compared with the best-fit
standard description of equation \eref{eq:RPD_SM_rate} with the addition
of radiative corrections by Bystritsky \etal \cite{Bys04}.  No
statistically significant deviations were observed.

The wider kinematic coverage of the new measurement enabled a much
improved analysis, including an independent determination of the
polar- and axial-vector form factors $F_V$ and $F_A$, and, for the first time, a
determination of the 4-momentum transfer $q^2=(p_e+p_\nu)^2$ dependence
of the polar vector amplitude $F_V$, \fref{fig:slope-2013}:
\begin{equation}
   F_V(q^2)=F_V(0)+aq^2 \qquad {\rm with} \qquad a=0.10\pm 0.06\,.
\end{equation}
The uncertainties on the slope parameter are wide enough to accommodate
the resonance chiral perturbation calculation of Mateu and Portol\'es
\cite{Mat07}.

Even after the inclusion of the lower-energy phase space regions, the
PIBETA \peiig\ decay data are strongly dominated by the $SD^+$
amplitude, leading to a stringent constraint on $F_V+F_A$, and a lax one
on $F_V-F_A$.  As in previous measurements, the probability of the
negative solution for $\gamma=F_A/F_V$ is disfavored by a large factor,
about 880.  The narrow elliptically shaped locus of best-fit values for
$F_A$ and $F_V$ is shown in \fref{fig:F_A-F_V-2013}.  The corresponding
equation of the major axis of the ellipse is given by
\begin{equation}
   F_A=(-1.0286 \cdot F_V + 0.03853) \pm 0.00014 \, .
      \label{eq:fa_fv_fit}
\end{equation}
We note that, for a fixed value of $F_V$, the uncertainty in $F_A$ has
been reduced 16-fold compared to the prior world average.  In a fully
unconstrained fit, the authors report
\begin{equation}
   F_V=0.0258 \pm 0.0017 \qquad {\rm and} \qquad 
   F_A=0.0117 \pm 0.0017\,.
\end{equation} 
The PIBETA result for $F_V$ represents a 5-fold improvement over the
prior average.  The correlated nature of the PIBETA form factor results
allows for a posteriori improvement in precision of $F_A$ as the
precision of the determination of the $\pi^0$ lifetime improves.

Finally, thanks to the experiment's broad kinematic coverage, the PIBETA
authors have reported a branching ratio
\begin{equation}
  B_{\pi_{e2\gamma}}(E_\gamma > 10\, {\rm MeV},
  \theta_{e\gamma} > 40^\circ) = 73.86(54) \times 10^{-8}\,,
\end{equation}
with more than an order of magnitude improved precision over the
previous result.

\subsection{Prospects for improvement}

The PEN experiment will add new data to the existing \peiig\ data set.
Since the PEN beam stopping rate is significantly lower than that used
in the 2004 PIBETA run, the impact on the statistical uncertainties will
not be great.  However, the cleaner running conditions of PEN present
greater access to the kinematic region $E_\gamma,E_e < 50$\,MeV,
previously strongly contaminated by muon decay background, and, thus,
holding the prospect for a better constraint of the $SD^-$ amplitude,
which, in turn, would lead to improved precision in the direct
determination of $F_V$.  These results will be forthcoming in the near
future.

\noindent 
\section{\texorpdfstring
     {\boldmath Pion beta decay: $\pi^+\to\pi^0 e^+\nu$ (\peiii)} 
     {Pion beta decay: pi+ to pi0 e+ nu  (pi-e3)}\label{sec:pi-beta} }

\subsection{General considerations: quark-lepton universality}
Unlike the $\pi_{e2(\gamma)}$ decays discussed in the preceding
sections, the extremely rare, ${\cal O}(10^{-8})$, pion beta decay is
not suppressed by any factor apart from the restricted phase space of
final states due to the small difference between the charged and neutral
pion masses, $\Delta=m_{\pm}-m_{0}$.  As a pure vector $0^-\to 0^-$
transition, it is completely analogous to the superallowed Fermi nuclear
beta decays.  In fact, it can be regarded as the simplest realization of
the latter, fully free of complications arising from nuclear structure
corrections.  Superallowed Fermi decays have historically led to the
formulation of the conserved vector current hypothesis, and have played
a critical role in testing the unitarity of the
Cabibbo-Kobayashi-Maskawa quark mixing matrix through evaluations of the
$V_{ud}$ element\cite{Cza04,Mar06,Har09}.

Within the framework of the \VmA\ theory of the weak interactions, the
pion beta decay rate can be expressed in terms of the leading-order
width $\Gamma_0$ and the radiative and loop corrections $\delta_\pi$ as
\cite{Kal64,Sir78}
\begin{equation}
     \Gamma = \Gamma_0(1+\delta_\pi)
         = \frac{G^2_F|V_{ud}|^2}{30\pi^3}\Delta^5 f(\epsilon,\Delta)
           \left(1-\frac{\Delta}{2m_+}\right)^{\!\!3}(1+\delta_\pi)\,,
           \label{eq:pb_theo_rate}
\end{equation}
where $G_F$ is the Fermi coupling constant and $\epsilon =
m_e^2/\Delta^2 \simeq 0.012375$.  Finally, up to leading order in
$\Delta^2/(m_++m_0)^2 \simeq 2.8 \times 10^{-4}$, the function
$f(\epsilon,\Delta)$ has the value of $f(\epsilon,\Delta) \simeq
0.94104$.  The overall uncertainty of the rate in \eref{eq:pb_theo_rate}
is dominated by two comparable contributions, one from the $\Delta^5$
factor, and the other from the $\delta_\pi$ radiative/loop corrections,
each uncertainty contribution being in the range of 0.05--0.1\% of the
full rate \cite{PDG12,Cir03,Mar06,Pas11}.  Thus, the pion beta decay
rate provides a direct means to determine the CKM matrix element
$|V_{ud}|$.  In fact, being free of nuclear structure corrections
present in superallowed nuclear beta decays, and free of tree-level
axial corrections present in neutron beta decay, \peiii\ decay offers
the theoretically cleanest path to measuring $V_{ud}$ and, hence,
testing quark-lepton universality.  However, the extremely low branching
ratio for the process has so far limited the experimental accuracy.

Experimentally, $\pi^+\to\pi^0e^+\nu_e$ decay is observed primarily
through detection of the final decay particles produced in the
near-instantaneous neutral pion decay ($\tau \simeq 8.5\times
10^{-17}$\,s).  The positron is generally harder to detect with precise
efficiency, except for $\pi^+$ decays in flight, since its kinetic
energy ranges from 0 to only 4\,MeV.  The key $\pi^0$ decay modes and
their branching ratios are:
\begin{eqnarray}
   \pi^0 \to & \gamma\gamma\,,  & \quad B \simeq 0.988\,, \\
             & \gamma e^+e^-\,, & \quad B \simeq 0.012\,. \nonumber
\end{eqnarray}
All experiments to date have focused on the $2\gamma$ channel rather
than the $\gamma e^+e^-$, Dalitz mode.  Due to the small charged to
neutral pion mass difference, the maximum kinetic energy of the $\pi^0$
is low, about 75\,keV.  Thanks to the correspondingly low $\pi^0$
velocity, the directions of the two emitted photons deviate from
$180^\circ$ by no more than $3.8^\circ$, on average by $\sim 3.2^\circ$.
Furthermore, the Doppler broadening of the photon energies results in a
narrow boxlike spectrum between $\sim 65.6$\,MeV and $\sim 69.4$\,MeV.
This kinematics provides a robust signal that is additionally separated
in time from the dominant prompt hadronic background events thanks to
the 26\,ns pion lifetime.

\subsection{Early measurements}

The first observation \cite{Dep62} of the decay at CERN in 1962 using a
stopped $\pi^+$ beam and a combination of lead glass and NaI(Tl)
detectors, was followed by a quick succession of early measurements at
CERN \cite{Dep63a}, Dubna (total absorption Pb glass \v{C}erenkov
counters) \cite{Dun64}, Columbia University \cite{Bar64}, Lawrence
Berkeley Lab \cite{Bac65} and Carnegie Tech (now Carnegie Mellon)
\cite{Ber65}.  The Berkeley, Columbia and Carnegie Tech experiments used
combinations of spark chambers and scintillation detectors.  The
five measurements achieved approximately 20\% uncertainties on the
branching ratio, based on samples of between 30 and 50 events each.

To date only three measurements of the $\pi^+\to\pi^0e^+\nu_e$ decay
branching ratio have been made with precision better than 10\%,
approximately one for every 20 years, which reflects the challenges of
the task.

\begin{figure}[htb]
  \centerline{\includegraphics[width=0.8\linewidth]
                   {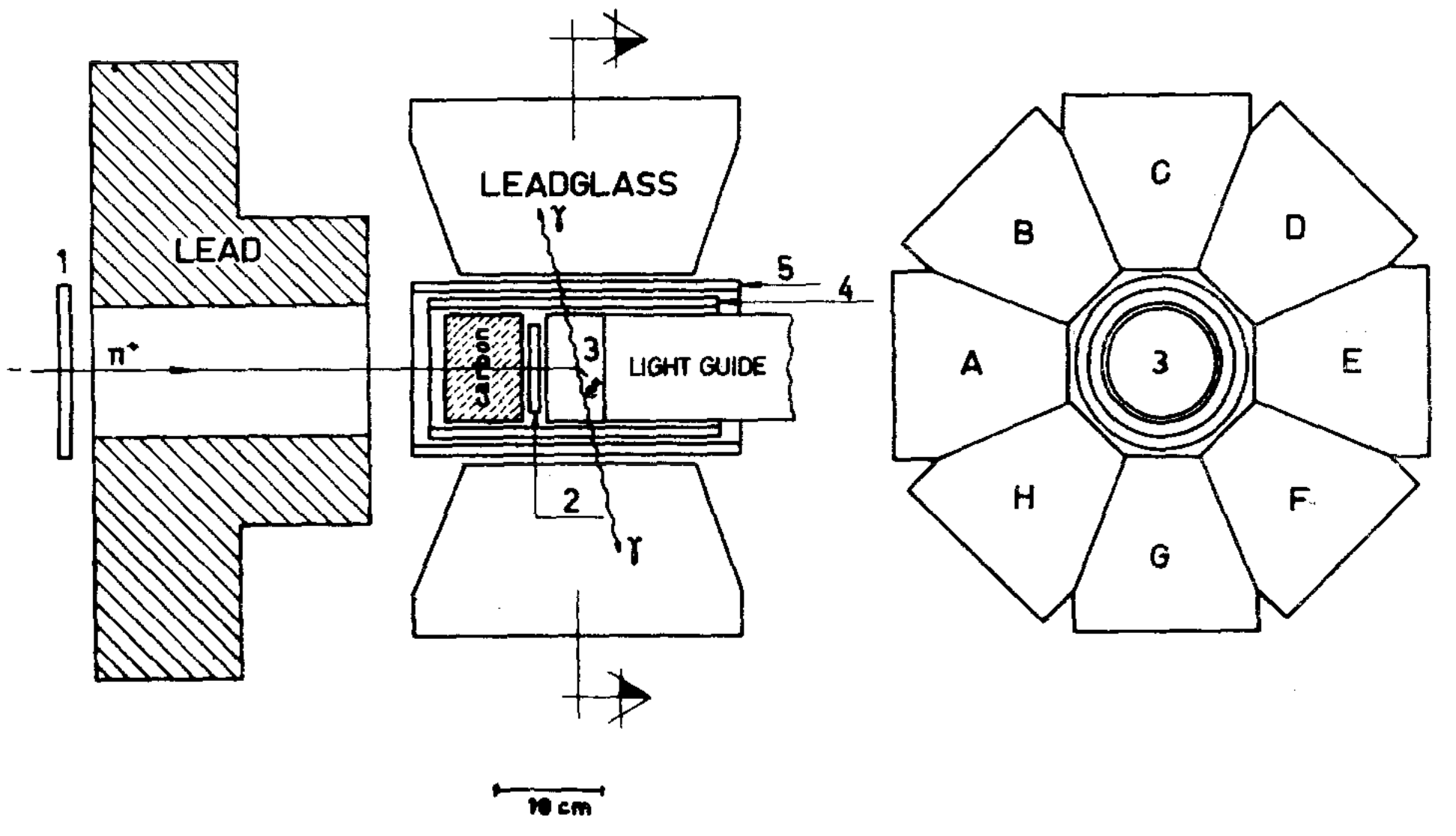}  }
  \caption{Schematic representation of the CERN pion beta decay
    apparatus of Depommier \etal\ used in their 1967 measurement
    \cite{Dep68}.  Plastic scintillator detectors 1, 2 and 3 define a
    beam $\pi^+$ stop.  Candidate pion beta decay events are recorded as
    narrow coincidences of a pair of electromagnetic showers in the lead
    glass detectors A through H, delayed with respect to the $\pi^+$
    stop pulse.  Detectors 4 and 5 veto charged particle events in the
    calorimeter array. For details see \cite{Dep68}. }
  \label{fig:cern_Pie3_apparatus}
\end{figure}

The CERN group of Depommier \etal \cite{Dep68} was first to break
the 10\% uncertainty level in 1967, using a lead glass \v{C}erenkov
detector array along with plastic scintillator detectors, as shown in
\fref{fig:cern_Pie3_apparatus}.
The CERN apparatus had an acceptance of $\sim\,$22.4\%, and enabled the
experimenters to record $332 \pm 10_{\rm bgd} \pm 23_{\rm stat} $ pion
beta decays from $\sim 1.5 \times 10^{11}$ pion stops in the target.
The reported branching ratio value was
\begin{equation}
  B(\pi_{e3})
    \equiv \frac{\Gamma(\pi^+\to\pi^0e^+\nu)}{\Gamma(\pi^+\to\mu^+\nu)}
    = \left(1.00\, ^{+0.08}_{-0.10} \right) \times 10^{-8}\,,
\end{equation}
in very good agreement with CVC theory predictions, but not precise
enough to test the radiative corrections at $\sim 4\%$.  The authors
increased the lower-side uncertainty of their result in order to account
for the possible influence of nuclear reaction products in their
measurement of the $\pi^0$ detection efficiency of the lead glass
detector array.

\begin{figure}[htb] 
  \centerline{\includegraphics[width=0.4\linewidth]
                   {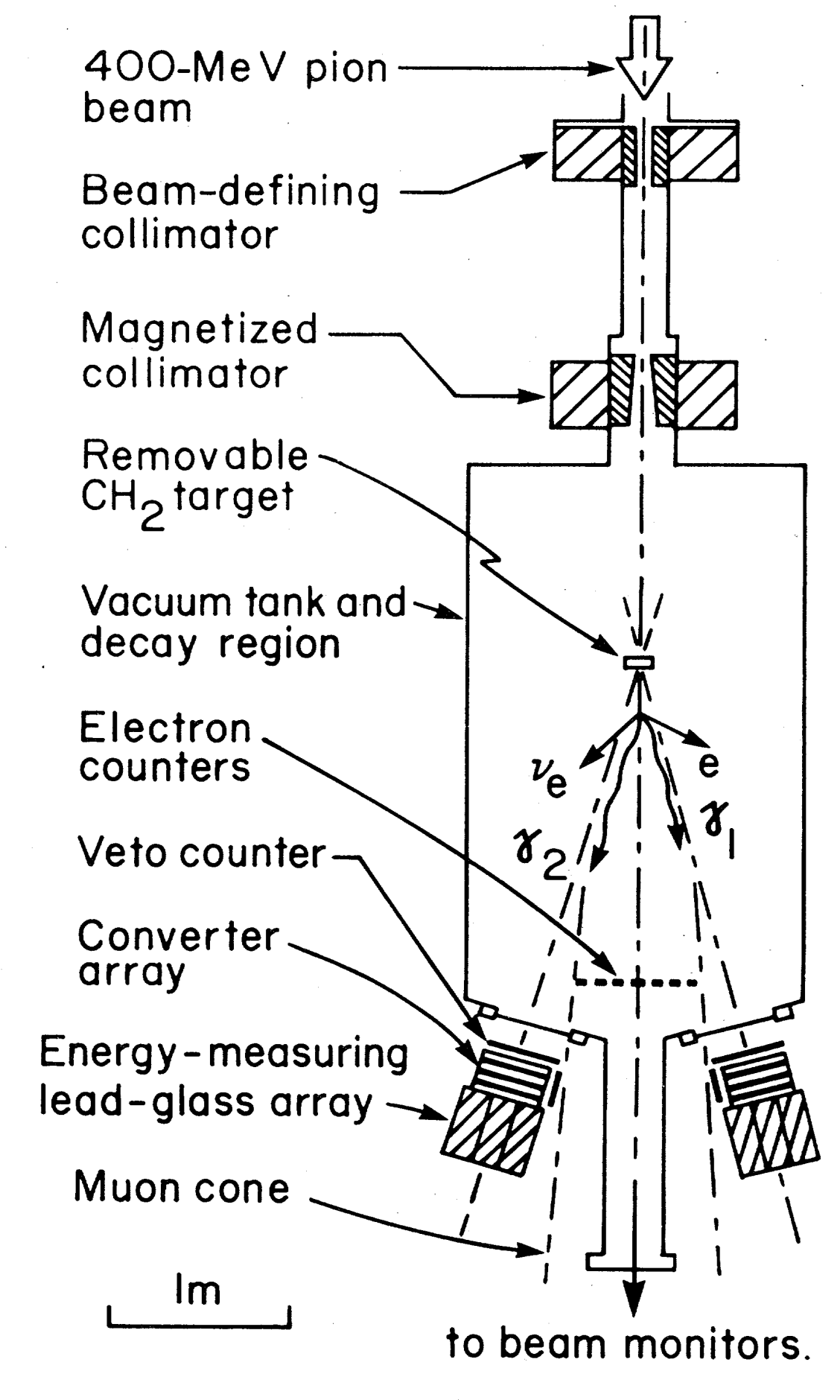}  }
  \caption{Schematic representation of the Temple--Los Alamos pion beta
    decay apparatus that McFarlane \etal\ used in their 1984
    measurement \cite{McF85}.  The removable CH$_2$ target was used for
    calibrations and removed during decay measurements.  For further
    details see \cite{McF85}.}
  \label{fig:lampf_Pie3_apparatus}
\end{figure}

In the early 1980's a Temple--Los Alamos group performed a radically
different measurement of the pion beta decay rate at the Los Alamos
Meson Physics Facility (LAMPF) \cite{McF85}.  Unlike previous
experiments which detected decays of $\pi^+$'s brought to rest in a
target, McFarlane and collaborators detected decays of an intense pion
beam in flight in the apparatus shown in \fref{fig:lampf_Pie3_apparatus}.
At $2\times 10^8$ pions/s and with a momentum of $522.1 \pm 0.8\,{\rm
MeV}/c$, the LAMPF pion beam was orders of magnitude more intense and an
order of magnitude higher in energy than in the previous experiments.
The decay volume was evacuated to $\sim 10^{-7}$\,torr in order to limit
the pion hadronic interactions with material in the path, primarily
single charge exchange reactions which would be hard to distinguish from
pion beta decay events.  Because of the low recoil in the $\pi^+$ rest
frame, the $\pi^0$ had transverse momentum of less than 5\,MeV/$c$, or
1\% of the longitudinal momentum.  The average $\pi^0$ momentum of about
505\,MeV/$c$ implied a short mean $\pi^0$ laboratory flight length of
about 100\,nm before decay.  The acceptance was defined by the
combination of an upstream collimator, minimum opening angle of the two
photons, and the solid angle of the active area of the two photon
detectors.  The latter were the two arms of the LAMPF $\pi^0$
spectrometer \cite{Bae80}, appropriately modified \cite{McF85} for the
pion beta decay measurement.  Each photon detector arm consisted of a
front plastic scintillator veto detector followed by three successive
lead glass converter layers (0.56 radiation length thick, each).  Each
converter layer was followed by two scintillation hodoscopes for
position and time measurement.  Finally, behind the forward crate
containing the converter-hodoscope layers was a rear crate with an array
of 15 lead glass blocks, forming a 14 radiation lengths thick
calorimeter.  The cross sectional area of the detector package as seen
by the photons was ${\rm 45\,cm\times 75\,cm}$.  The arms were well
matched to detecting the final-state photons that ranged between 175 and
350\,MeV in energy.  At $\sigma_t \simeq 250$\,ps per arm, the time
resolution was excellent for such large detectors.  At $\sigma_E \simeq
1.53\sqrt{E}$ (in MeV), the energy resolution was adequate for the task.

This measurement apparatus has several advantages over the stopped decay
technique.  First, the hadronic background can be effectively eliminated
in an evacuated decay region.  Second, the high pion beam momentum
focuses the decay photons into a solid angle significantly smaller than
$4\pi$\,sr.  Finally, in principle, the measurement can be done with
both charged pion states, although in practice the intensity of $\pi^-$
beams is significantly lower than that of $\pi^+$.  

However, there are also significant disadvantages compared to the
stopped decay measurement.  The proper time that beam pions spend in the
decay region is very short, about $10^{-3}$ pion lifetimes, requiring
enormously higher beam intensities to achieve high statistical precision
for the rare pion beta decays.  Normalization, i.e., counting the number
of pions passing through the apparatus is very challenging, as the
readily available processes such as the $\pi_{\mu2}$ decay have
radically different kinematics and detection systematics, and require
separate detectors.  Finally, detection efficiency (including detector
acceptance), as well as trigger and event selection efficiency must be
determined in absolute terms with high precision.

Given all the above considerations, the Temple--LAMPF group reported the
experimental partial rate, $1/\tau_{\pi\beta}$, and branching ratio,
$B_{\pi\beta}$: 
\begin{equation}
  \frac{1}{\tau_{\pi\beta}} = (0.394 \pm 0.015)\,{\rm s}^{-1}
  \qquad {\rm or} \qquad
  B_{\pi\beta} = (1.026 \pm 0.039)\times 10^{-8}\,,
\end{equation}
based on $1124 \pm 36$ pion beta decay events, and $2.14 \times 10^{14}$
beam pions.  The overall 3.8\% uncertainty breaks down to 3.1\%
statistical and 2.0\% systematic contributions.  The largest systematic
uncertainty contributions arose from the number of beam pions, time in
decay region, and the various efficiencies.  The authors also extracted
a new value of $0.964\pm 0.019$ for the cosine of the Cabibbo angle, or
$V_{ud}$.  While not as precise as the contemporary values of $V_{ud}$
derived from superallowed Fermi nuclear beta decays, it was consistent
with them.

\subsection{The PIBETA experiment}

\begin{figure}[htb]
   \centerline{\includegraphics[width=0.67\linewidth]{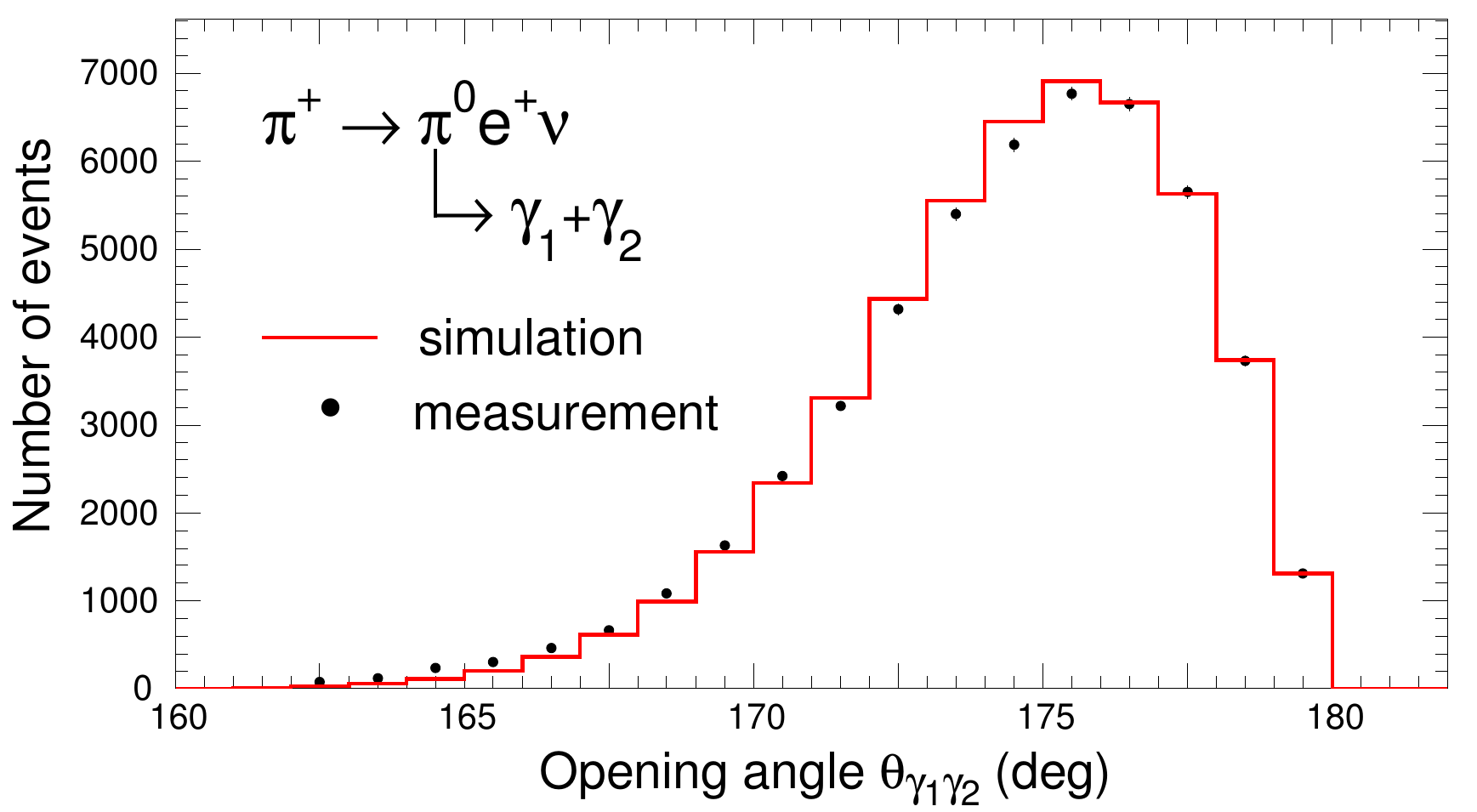}}
   \caption{Distribution of $\gamma_1-\gamma_2$ photon opening angle in
     the PIBETA experiment for the pion beta decay event sample.  Data
     points reflect measurements, while the red histogram indicates
     results of a realistic GEANT3 Monte Carlo simulation including the
     full effects of the PIBETA detector resolution.  Radiative
     \peiig\ decay events, if misidentified as \peiii\ events through
     positron annihilation in flight, have a significantly different
     $\theta_{\gamma1\gamma2}$ signature.}
   \label{fig:pb_ang12}
\end{figure}

\begin{figure}[htb]
    \centerline{\includegraphics[width=0.7\linewidth]{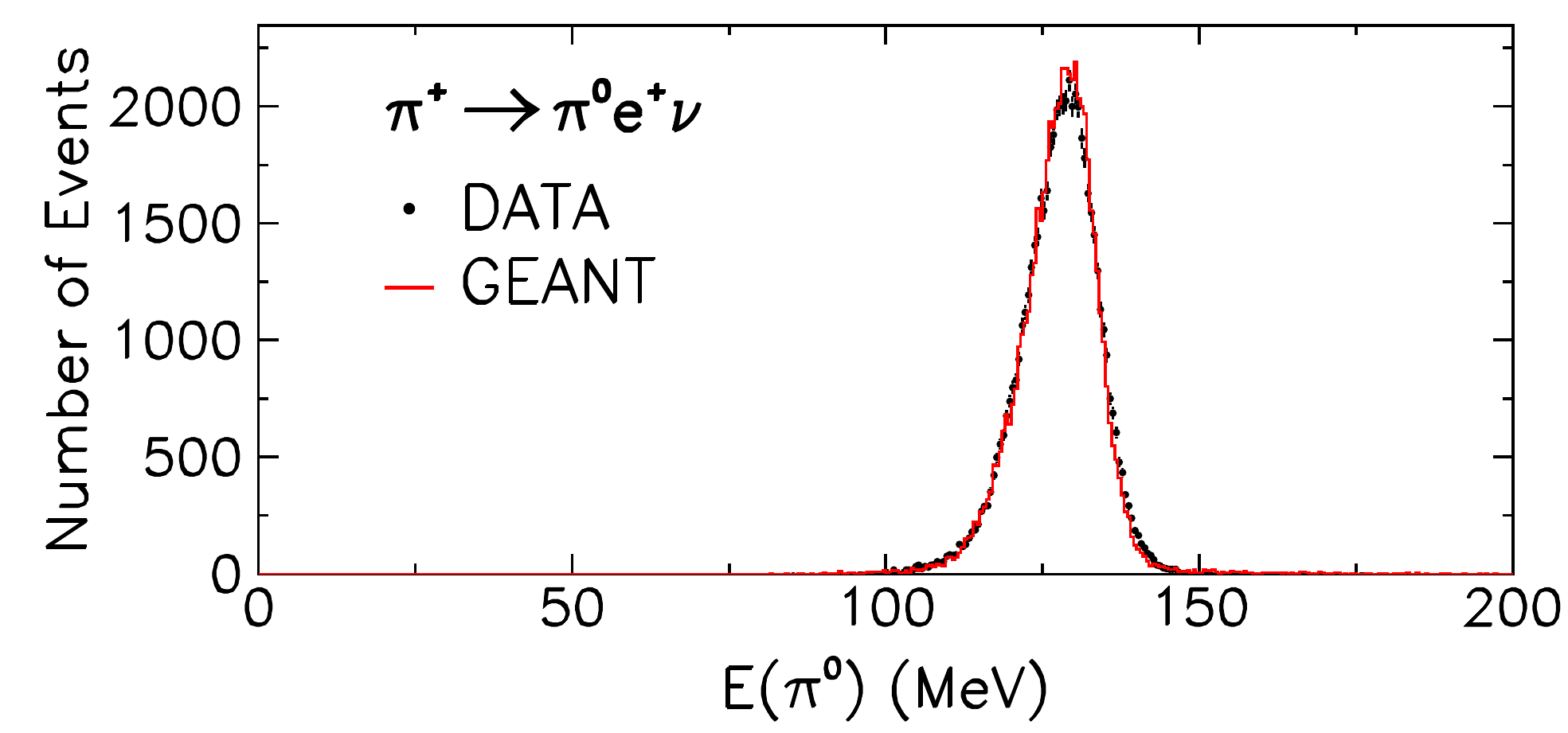}} 
   \caption{Distribution of the $\pi^0$ energy in the PIBETA experiment.
     Data points reflect measurements, while the red histogram indicates
     results of a realistic GEANT3 Monte Carlo simulation including the
     full effects of the PIBETA detector resolution. }
   \label{fig:pb_energ}
\end{figure}

\begin{figure}[htb]
    \centerline{\includegraphics[width=0.7\linewidth]{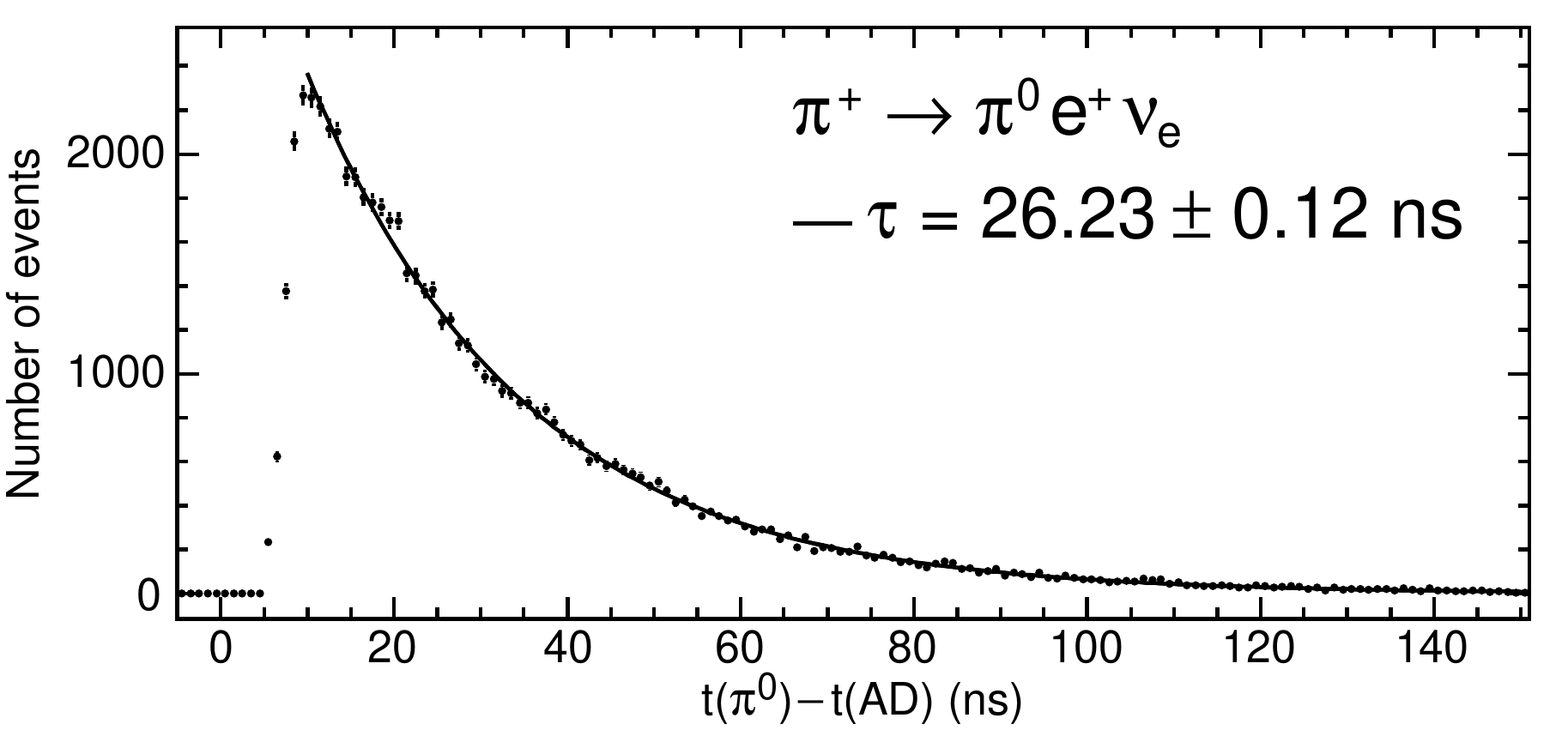}}
   \caption{Distribution of the $\pi^0$ event times with respect to the
     time $t=0$ of the active degrader (AD) signal, reflecting the
     beam pion stop time, in the PIBETA experiment.  The curve is a
     best-fit exponential decay with the decay parameter in good
     agreement with 26.03\,ns, the pion lifetime.  The time range of
     $\pm 5$\,ns around $t=0$, the pion stop time, was vetoed in the
     decay event trigger in order to suppress prompt hadronic
     backgrounds which had a $\sim 10^4$ times higher rate than
     the \peiii\ events.  A prescaled sample of the prompt hadronic
     events was collected for systematic studies and
     calibrations. } \label{fig:pb_time}
\end{figure}

The primary goal of the PIBETA experiment conducted at PSI and
introduced in sections \ref{sec:pi_e2} and \ref{sec:rpd}, was to improve
by an order of magnitude the existing experimental precision of the pion
beta decay branching ratio.  The apparatus is described schematically in
\fref{fig:PEN_det}, while the central detector region is shown in
\fref{fig:cent_det}.  A $\sim$114\,MeV/$c$ pion beam of the PSI $\pi$E1
beamline was focused onto the segmented nine-element active target which
stopped an average of $0.8 - 1 \times 10^6$ pions/s.  The PIBETA
approach was essentially similar to the 1967 CERN experiment of
Depommier \etal \cite{Dep68}, but with larger acceptance, much improved
energy resolution of the electromagnetic shower calorimeter, MWPC
tracking of charged particles between the central region and the
calorimeter, a more intense pion beam, and faster electronics and data
acquisition system.  The key detector response functions to pion beta
decay events are shown in figures \ref{fig:pb_ang12}, \ref{fig:pb_energ}
and \ref{fig:pb_time}.
The corresponding spectra for the normalizing \peii\ decay, and for the
\peiig\ decay which can contribute background events for the pion beta
decay, are shown in sections \ref{sec:pen} and \ref{sec:pibeta-rpd}.
Since a subset of pion beta decays can also be misidentified as
radiative \peiig\ decays, it was essential that all three processes be
simultaneously accounted for and understood at the few parts per
thousand level.  In fact, ordinary (``Michel'') and radiative muon
decays, though not discussed here, were included as well in the
comprehensive PIBETA analysis, since they comprise the majority of
actual events occurring in the apparatus, and dominate the accidental
background to all pion decay channels.
Figures \ref{fig:pb_ang12}\,--\,\ref{fig:pb_time} demonstrate that the
pion beta decay event sample was clean and well described in terms of
the relevant instrumental resolutions.

During three runs in 1999, 2000 and 2001, the PIBETA collaboration
acquired over 64,000 pion beta decay events, which led to an improvement
of the experimental precision of the \peiii\ branching ratio of about an
order of magnitude \cite{Poc04}.  The PIBETA result was evaluated in two
ways, (a) normalizing to the average experimental \peii\ branching ratio
of \eref{eq:pi_e2_avg} $R_{\e/\mu}^{\pi}= (1.230\pm 0.004)\times
10^{-4}$ (``exp.\ norm.''), and (b) normalizing to the established
theoretical value of \eref{eq:pi_e2_full_SM}, $R_{\e/\mu}^{\pi}=
(1.2352\pm 0.0005)\times 10^{-4}$ (``theo. norm.'').  The resulting
values were:
\begin{eqnarray}
   B^{\rm exp.\ norm.}_{\pi\beta}  
        = \left[1.036 \pm 0.004({\rm stat}) \pm 0.004({\rm syst})
           \pm 0.003(\pi_{e2})\right]  \times 10^{-8}, \\
   B^{\rm theo.\ norm.}_{\pi\beta}  
        = \left[1.040 \pm 0.004({\rm stat}) \pm 0.004({\rm syst})\right]
            \times 10^{-8}\,.
\end{eqnarray}
The external experimental uncertainty included the branching ratios
$R_{\e/\mu}^{\rm\pi\ exp}$, $B_{\pi^0\to\gamma\gamma}^{\rm exp}$ and
that of $\tau_{\pi^+}^{\rm exp}$, the pion lifetime, and were dominated
by the former.  The combined internal systematic uncertainty was
dominated by the uncertainties related to the ratio of the ``pion gate''
live fractions for the \peiii\ and \peii\ processes, and the number of
the \peii\ events used for normalization, followed by the ratio of
acceptances for the same two processes.

This result represents the most stringent test of CVC and Cabibbo
universality in a meson, as well as of the proper treatment of the
radiative corrections, which, combined, predict $B_{\pi\beta}^{\rm SM} =
(1.038 - 1.040)\times 10^{-8}$ at 90\% confidence limits \cite{PDG04}.
Excluding radiative corrections gives the range $B_{\pi\beta}^{\rm SM} =
(1.005 - 1.007) \times 10^{-8}$, a $>4\sigma$ discrepancy from the
PIBETA result.  The PIBETA collaborators also extracted
\begin{eqnarray}
   V_{ud}^{\rm\pi\beta/exp.\ norm.} = 0.9728 \pm 0.0030\,,\ {\rm and} \\
   V_{ud}^{\rm\pi\beta/theo.\ norm.} = 0.9748 \pm 0.0025\,.
\end{eqnarray}
These values are in excellent agreement with the PDG recommended value
for $V_{ud}$, although five times less precise.

Following in the same vein, it is tempting to go one step further and
turn the PIBETA determination of the $\pi_{e3}$ branching ratio around,
using it instead to evaluate $R_{e/\mu}^\pi$.  This is accomplished by
fixing $V_{ud}$ to its extraordinarily precise PDG 2013 recommended
value of $0.97425 \pm 0.00022$ and adjusting $R_{e/\mu}^\pi$ until the
extracted value of $V_{ud}^{\pi\beta}$ agrees.  This intriguing exercise
yields
\begin{equation}
   (R_{e/\mu}^\pi)^{\rm PIBETA} = (1.2366 \pm 0.0064) \times 10^{-4}\,,
\end{equation}
in good agreement with direct measurements reviewed in \sref{sec:pi_e2}.
Appropriately averaging this value of $R_{e/\mu}^\pi$ with those listed
in \tref{tab:pe2_BRs}, one obtains a slightly higher value than the
current PDG average: $(R_{e/\mu}^\pi)^{\rm new\ avg} = (1.2317 \pm
0.0031) \times 10^{-4}$.

Given the advantage in lower theoretical uncertainties compared to
nuclear decays, there was every incentive to pursue a higher precision
result in the pion beta decay rate or branching ratio.  However, the
urgency of a further improvement was considerably reduced by the
Brookhaven National Laboratory experiment BNL E865 result\cite{She03}
and the subsequent renormalization of $V_{us}$ that removed a
longstanding 2--3$\sigma$ shortfall in CKM matrix unitarity.  In light
of the considerable experimental challenges that an improved measurement
would pose, the present full agreement of available data on $V_{ud}$
with CKM unitarity means that a major new project on improving the
precision of pion beta decay is not currently planned.  We note, though,
that the status of neutron beta decay is in considerable flux, with
major discrepancies in the available data sets on both the beta
asymmetry and the neutron lifetime \cite{Bae14,You14}.  At present there
are a number of projects addressing these deficiencies.  While not as
theoretically clean as the pion beta decay, neutron decay is free of the
nuclear structure and Coulomb corrections that affect nuclear beta
decays, including the superallowed Fermi $0^+\to 0^+$ transitions
\cite{Hol14}. 

\section{Summary}

A close look at the recent record of study of the decays of the charged
pion reveals a great deal of activity and continued strong relevance.
The extraordinary precision of the theoretical description of the
$\pi_{l2}$, $\pi_{l2\gamma}$, $\pi_{l2e^+e^-}$ and $\pi_{e3}$ decays
remains unmatched by the available experimental results.  There have
been impressive advances in the precision of the \peiii\ and
\peiig\ decays in the past decade.  As of this writing, we are on the
verge of a significant improvement of the \peii\ decay precision, as the
PEN and PiENu experiments complete their analyses.  Even subsequent to
that development, there will remain considerable room for improvement
of experimental precision with high payoff in terms of limits on physics
not included in the present Standard Model.  This work remains relevant
and complementary to the direct searches on the energy frontier
currently under way at particle colliders, with considerable theoretical
significance.

\section*{Acknowledgments}

This work has been supported by grants from the US National Science
Foundation (most recently PHY-1307328) and the Paul Scherrer Institute.

\end{document}